\documentclass[twocolumn]{pasj01}

\usepackage[switch,mathlines]{lineno}

\Received{2023/07/31}
\Accepted{2023/09/23}
\Published{}

\usepackage{natbib}
\usepackage{graphicx}	
\usepackage{color}
\usepackage{tabularx}
\usepackage{dcolumn}
\usepackage{booktabs}
\usepackage{soul}

\newcommand{\sun}{\odot}
\newcommand{\rmxaa}{RMxAA}
\newcommand{\nodata}{$\cdots$}

\newcommand{\oi}{[O\,{\sc i}]}
\newcommand{\oii}{[O\,{\sc ii}]}
\newcommand{\oiii}{[O\,{\sc iii}]}
\newcommand{\oiv}{[O\,{\sc iv}]}

\newcommand{\NI}{[N\,{\sc i}]}
\newcommand{\nii}{[N\,{\sc ii}]}

\newcommand{\sii}{[S\,{\sc ii}]}
\newcommand{\siii}{[S\,{\sc iii}]}
\newcommand{\siv}{[S\,{\sc iv}]}

\newcommand{\hei}{He\,{\sc i}}
\newcommand{\heii}{He\,{\sc ii}}

\newcommand{\neii}{[Ne\,{\sc ii}]}
\newcommand{\neiii}{[Ne\,{\sc iii}]}

\newcommand{\ariii}{[Ar\,{\sc iii}]}
\newcommand{\ariv}{[Ar\,{\sc iv}]}

\newcommand{\cii}{C\,{\sc ii}}

\newcommand{\civ}{C\,{\sc iv}}

\newcommand{\ha}{H$\alpha$}
\newcommand{\hb}{H$\beta$}

\newcommand{\hi}{H\,{\sc i}}

\newcommand{\kms}{km\,s$^{-1}$}
\newcommand{\te}{$T_{\rm e}$}
\newcommand{\Ne}{$n_{\rm e}$}
\newcommand{\ergs}{erg\,s$^{-1}$\,cm$^{-2}$\,{\AA}$^{-1}$}
\newcommand{\ergsF}{erg\,s$^{-1}$\,cm$^{-2}$}
\newcommand{\ergsm}{erg\,s$^{-1}$\,cm$^{-2}$\,{\micron}$^{-1}$}

\begin{document} 

\title{ 
Seimei KOOLS-IFU mapping of the gas and dust distributions in Galactic PNe: Unveiling the origin and evolution of  Galactic halo PN H4-1}

\author{Masaaki \textsc{Otsuka}\altaffilmark{1}
\thanks{Corresponding author}
}
\altaffiltext{1}{Okayama Observatory, Kyoto University, Honjo, Kamogata, Asakuchi, Okayama 719-0232, Japan}
\email{otsuka@kusastro.kyoto-u.ac.jp}

\author{Toshiya \textsc{Ueta},\altaffilmark{2}}
\altaffiltext{2}{Department of Physics and Astronomy, University of Denver, 2112 E Wesley Ave., Denver, CO 80208, USA}
\email{toshiya.ueta@du.edu}

\author{Akito \textsc{Tajitsu}\altaffilmark{3}}
\altaffiltext{3}{Subaru Telescope Okayama Branch Office, 
National Astronomical Observatory of Japan, National Institutes of Natural Sciences, 
3037-5 Honjo, Kamogata, Asakuchi, 
Okayama 719-0232, Japan}
\email{akito.tajitsu@nao.ac.jp}


\maketitle

\begin{abstract}
H4-1 is a planetary nebula (PN) located in the Galactic halo, and is notably carbon-rich and one of the most metal-deficient PNe in the Milky Way. 
To unveil its progenitor evolution through the accurate measurement of the gas mass, we conducted a comprehensive investigation of H4-1, using the newly obtained Seimei/KOOLS-IFU spectra and multiwavelength spectro-photometry data. The emission line images generated from the KOOLS-IFU datacube successfully resolve the ellipsoidal nebula and the equatorial flattened disk that are frequently seen in bipolar PNe evolved from massive progenitors. By a fully data-driven method, we directly derived the seven elemental abundances, the gas-to-dust mass ratio, and the gas and dust masses based on our own distance scale. By comparing the observed quantities with both the photoionization model and the binary nucleosynthesis model, we conclude that the progenitors of an initial mass of 1.87\,M$_{\sun}$ and 0.82\,M$_{\sun}$ are second generation stars formed $\sim4$\,Gyrs after the Big Bang, and underwent mass-transfers, binary merger, and ultimately evolved into a PN showing unique chemical abundances. Our binary model successfully reproduces the observed abundances and also explains evolutionary time scale of H4-1. 
\end{abstract}


\section{Introduction}
\label{S-intro}

\begin{table*}
\caption{Observed nebular elemental abundances of the halo PNe.}
\begin{tabularx}{\textwidth}{@{\extracolsep{\fill}}lcccccccccl}
\midrule
PN & He & C (CEL) & N & O & Ne & S & Cl & Ar &Fe & Ref. \\ 
\midrule
H4-1 & 11.02 & 8.86 & 7.74 & 8.17 & 6.42 & 5.14 & 3.88 & 4.61 &\nodata  &(1), (2)\\ 
K648 in M15 & 11.02 & 8.97 & 6.36 & 7.73 & 7.44 & 5.40 & 3.58 & 4.60 &5.02& (3) \\ 
BoBn1 & 11.07 & 9.02 & 8.03 & 7.74 & 7.96 & 5.32 & 3.39 & 4.33 &5.08  &(4) \\ 
Average & 11.04 & 8.95 & 7.38 & 7.88 & 7.27 & 5.29 & 3.62 & 4.51 &5.05 & \\
\midrule
DdDm1 & 11.01 & 7.02 & 7.36 & 8.07 & 7.45 & 6.35 & 4.61 & 5.79 &6.58 &(5) \\ 
NGC2242 & 11.00 & 8.39 & 7.72 & 8.03 & 7.80 & 6.33 & 4.45 & 5.89 &\nodata &(6),(7)\\
NGC4361 & 11.03 & 8.13 & 7.41 & 7.82 & 7.54 &\nodata  &\nodata  & 6.01 &\nodata&(7)\\ 
PN G135.9+55.9 & 10.95 & 7.84 & 7.15 & 6.82 & 6.83 & 5.50 & \nodata & 4.50 &\nodata& (8)\\ 
PRMG1 & 10.96 &\nodata  &\nodata  & 8.10 & 7.50 &\nodata  &\nodata  & 5.80 &\nodata &(9) \\ 
PRMT1 & 11,03 & 7.60 & 8.00 & 8.30 & 7.90 &\nodata  &\nodata  & 6.30 &\nodata &(10)\\ 
M2-29 in M22 & 10.97 & \nodata & 6.98 & 7.31 & 6.72 & 5.91 &  & 5.26 &\nodata &(11)\\ 
JaFu1 in Pal6 & 11.15 & \nodata & 7.97 & 8.49 & 8.00 & 6.62 &\nodata  &\nodata  &\nodata&  (12)\\ 
JaFu2 in NGC6441 & 11.06 & \nodata & 7.96 & 7.73 & 6.79 & 6.77 &\nodata  & 5.46 &\nodata& (12) \\ 
Average & 11.02 & 7.80 & 7.57 & 7.85 & 7.39 & 6.25 & 4.53 & 5.63 &6.58&  \\ 
\midrule
Sun & 10.93 & 8.43 & 7.83 & 8.69 & 7.93 & 7.12 & 5.50 & 6.40 &7.50  &(13)\\ 
\midrule
\end{tabularx}
\begin{tabnote}
Note -- The C abundances are derived from collisional excitation lines (CELs). The S abundance in NGC2242 is calculated by us using the data of \cite{2003PASP..115...80K}. 
There are no reports of the nebular elemental abundances of GJJC1 in M22. \\
References -- (1) This work; 
(2) \cite{Otsuka:2013aa}; 
(3) \citet{2015ApJS..217...22O}; 
(4) \cite{2010ApJ...723..658O};
(5) \cite{2009ApJ...705..509O};
(6) \cite{2003PASP..115...80K};
(7) \cite{1990A&A...233..540T};
(8)  \cite{2010A&A...511A..44S};
(9) \cite{1989RMxAA..17...25P};
(10) \cite{1990A&A...237..454P}; 
(11) \cite{1991PASP..103..865P};
(12) \cite{1997AJ....114.2611J};
(13) \cite{Asplund:2009aa}.
\end{tabnote}
\label{T-halopn}
\end{table*}

Planetary nebulae (PNe) are the final stage of evolution of low to intermediate mass stars with initial mass of $1-8$\,M$_{\sun}$ \citep[e.g.,][]{2000oepn.book.....K}. 
In our Galaxy, more than 3000 PNe have been identified so far \citep{2016JPhCS.728c2008P}, 
mostly belong to the Galactic disk. However,
there are also 13 halo PNe\footnote{\cite{2015ApJS..217...22O} reported 14 objects 
as halo PNe by following \cite{2007A&A...467.1249P} reported that SaSt2-3 is a new halo PN. 
Later, \cite{2019MNRAS.482.2354O} concluded that this PN is not halo but thin disk population.}, 
comprising eight field PNe in the Galactic halo 
and five members of globular clusters. 

Halo PNe are classified into type-IV on the basis of kinematical/chemical characteristics which was originally proposed by 
\citet{1978IAUS...76..215P} and later redefined by \citet{2007A&A...475..217Q}.
Halo PNe have low metallicity, with the average [Ar/H] of $-1.33$ among 12 halo PNe (table\,\ref{T-halopn}).
Despite their scarcity, halo PNe have been considered special because 
(i) they were formed in the early stages of Galaxy formation, and
(ii) they allow us to directly measure both chemical abundances and core masses of the central star after the asymptotic giant branch (AGB) star nucleosynthesis. 
To better understand halo PNe, we have investigated physical properties of 
K648 in the globular cluster M15 \citep{2015ApJS..217...22O}, BoBn1 \citep{2008ApJ...682L.105O,2010ApJ...723..658O}, 
DdDm1 \citep{2009ApJ...705..509O}, and H4-1 \citep{2003PASP..115...67O,Otsuka:2013aa}.
In this paper, we discuss the origin and evolution of H4-1.

H4-1 is one of the most metal-poor ($\mathrm{[Ar/H]} = -1.78 \pm 0.15$, corresponding to the metallicity of $Z\sim10^{-4}$) and 
carbon-rich ($\mathrm{C/O} = 4.90$) PNe in the Milky Way, and hence, 
is of special interest along with K648 and BoBn1, 
which share similar chemical properties (table\,\ref{T-halopn}). 
Also, H4-1 is remarkably $r$-process rich \citep[$\mathrm{[Xe/H]} > +0.51$;][]{Otsuka:2013aa}. 
Given the low metallicity, H4-1 is believed to have formed $\sim10-13$\,Gyrs ago.
However, our understanding of its evolution remains incomplete. 
More specifically, there have been longstanding unresolved issues on its slow evolutionary time scale and high C abundance. 
Based on the isochron fitting to the main-sequence turnoff point of M15 \citep[of $Z=10^{-4}$,][]{1985ApJS...58..225F}, H4-1 was considered to be of 0.8\,M$_{\sun}$ initial mass. 
Post-AGB evolutionary models of a 0.85\,M$_{\sun}$ and $Z=10^{-4}$ star by \citet{2016A&A...588A..25M} suggest that such stars would take $\gtrsim 15000$\,yrs to increase the effective temperature ($T_{\rm eff}$) from $\sim5000$\,K \citep[when mass loss is considered to be terminated;][]{1983ApJ...272..708S} to $\sim30000$\,K (when ionization of the circumstellar nebula begins). 
In addition, such stars are expected to spend $>60000$\,yrs to reach the current $T_{\rm eff}$ of $\sim120000$\,K from $T_{\rm eff}$ 
of $\sim5000$\,K (table\,\ref{T-dis}). 
It is highly unlikely to visually identify such extremely slow-evolving PNe. 
However, why can we still observe H4-1?

According to the stellar evolution models, third dredge-up (TDU) is required 
during the thermal pulse (TP) AGB phase in order for the low-intermediate mass to become carbon-rich.
The lower mass limit for TDU to happen is dependent on model assumptions; as far as we know, 
it is about 0.9\,M$_{\sun}$ \citep[e.g.][]{2012ApJ...747....2L}, which is more massive than an expected initial mass of H4-1. 
If so, how could H4-1 become C-rich?

Is H4-1 really evolved from a single star of about $0.8-0.9$\,M$_{\sun}$ initial mass? 
\citet{Otsuka:2013aa} found out that the [C/O] and [$s$-element/$r$-element] 
ratios in H4-1 are similar to those in 
carbon-enhanced metal-poor (CEMP) stars \citep[e.g.,][]{2005ARA&A..43..531B} which are believed to originate from binary systems.
Following the classification by \citet{2011MNRAS.412..843S} for metal-poor stars, H4-1 is CEMP-$r/s$ or CEMP-no \citep{Otsuka:2013aa}. 
Binary mass-transfer and merger would drastically shorten the evolutionary timescale, enabling an explanation of the evolutionary timescale. As such binary evolution would increase up to the mass required for TDUs, we would explain the C-richness of H4-1. 
Consequently, binary evolution would allow us to explain both the evolutionary timescale and 
the carbon enhancement without any contradictions. 
Indeed, there is a recent report on the detection of molecular hydrogen emission lines, 
which implies that H4-1 originated from a massive star \citep{tajitsu:2023aa}. 
Hence, we need to seek more conclusive evidence of its binary origin based on observational quantities.

One way to determine the origin and evolution of H4-1 is to estimate the total mass of the circumstellar matter. \citet{2016A&A...588A..25M} predicts that a single star of initially $0.85\,\mathrm{M}_{\sun}$ with $Z=10^{-4}$ would become a $0.53\,\mathrm{M}_{\sun}$ white dwarf. That is, such a star would eject 
the circumstellar mass of $0.32\,\mathrm{M}_{\sun}$ during its evolution. 
If the progenitor is a binary system consisting of a $\sim0.8\,\mathrm{M}_{\sun}$ star and a more massive star, 
a greater amount of mass would have to be ejected. 
The origin and evolution of H4-1 and its anomalous abundance pattern will also be important to understand the nature of CEMP stars.

To accurately estimate the total mass of the circumstellar matter, and subsequently, the mass of the progenitor system, an accurate distance estimate is critical. However, the distance toward H4-1 has not been 
fully established (see subsection\,\ref{S-dist}). The distance to H4-1 appears too far to be measured even with {\it Gaia} \citep{2021A&A...656A..51G}. 
Therefore, we undertake the distance determination on our own,  
which requires multi-wavelength spectro-photometry data and post-AGB evolutionary tracks.

This paper is organized as follows.  
In sections\,\ref{S-data} and \ref{S-image}, we explain a new set of data obtained with our own Seimei/KOOLS-IFU observations, 
the adopted archival data, and how the data are reduced, before presenting the obtained spectrum and emission line images. 
Then, 
a fully data-driven plasma diagnostics with self-consistent dust extinction correction is performed in section\,\ref{S-ext},
followed by a detailed abundance analysis in section\,\ref{S:abund}. 
The gas and dust masses and the gas-to-dust mass ratio are calculated in section\,\ref{S-gdr}. 
We have discussions on the evolution of the progenitor in section\,\ref{S-dis}, 
and summarize the current work in section\,\ref{S-sum}.

\begin{figure}
\begin{center}
\includegraphics[width=\columnwidth]{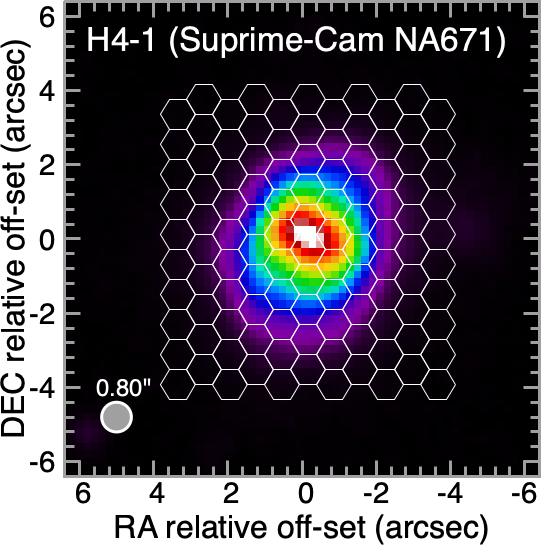}
\end{center}
\caption{
One of the dither positions of the packed fiber array in KOOLS-IFU observations overlaid on the Subaru 
8.2-m/Suprime-Cam image presented in \citet{tajitsu:2023aa}. 
The intensity scale ranges between the minimum and the maximum.
The spatial resolution of the image is indicated by the gray circle. 
Each hexagon indicates the position and dimension of the 110 fibers. 
\label{F:koolsfv}} 
\end{figure}

\section{Dataset and reduction}
\label{S-data}

We explain the used dataset and the data reduction below. 
The logs of the spectroscopic observations are summarized in table\,\ref{T-obslog}.

\subsection{Seimei/KOOLS-IFU optical spectra} 
\label{S:kools}

We performed the mapping observation (figure\,\ref{F:koolsfv}) using the 
Kyoto Okayama Optical Low-dispersion Spectrograph with optical-fiber integral field unit 
\citep[KOOLS-IFU;][]{2019PASJ...71..102M} attached to one of the two Nasmyth foci of 
the Kyoto University Seimei 3.8-m telescope \citep{2020PASJ...72...48K}. 

The observations were carried out under clear and stable sky conditions on three nights in early 2023. 
The seeing 
was $\sim2{\arcsec}$ in $R$-band at full width at the half maximum (FWHM). The data were taken under the airmass of 
$1.10-1.00$ in the VPH-blue and $1.06-1.01$ in the VPH-red settings. We took 
$12\times600$\,sec (2023 March) and $4\times1200$\,sec (2023 April) 
in the blue and $8\times900$\,sec in the red observations, 
respectively. 
The logs of the spectroscopic observations are summarized in table\,\ref{T-obslog}.

For the background sky subtraction, the off-source frames were taken every three science frames. 
To perform flux calibration and remove telluric absorption, 
we observed Hz44 every 30 min. 
This standard star was also used to evaluate light loss in the KOOLS-IFU field and synthesize a point-spread function (PSF).
We also took instrumental flat, twilight sky, and Hg/Ne/Xe arc frames as the baseline calibration.

We reduced the data using our own codes and {\sc IRAF} \citep{1986SPIE..627..733T} by following the procedure elucidated by \citet{2022MNRAS.511.4774O}. 
Wavelength calibration was performed using over 40 Hg/Ne/Xe lines that 
cover the full range of effective wavelengths. 
We adopted a constant plate scale of 2.5\,{\AA}. 
The resultant RMS errors in wavelength determination were 0.11\,{\AA} in the VPH-blue 
and 0.13\,{\AA} in the VPH-red, respectively. 
The average spectral resolution ($R$, defined as $\lambda$/Gaussian FWHM) measured from the Hg/Ne/Xe 
lines was $\sim500$ and $\sim800$ in the VPH-blue and VPH-red spectra, respectively. 

We obtained an RA-DEC-wavelength datacube using our codes and image reconstruction based 
on the drizzle technique. We corrected the spatial displacements due to atmospheric dispersion 
at each wavelength by following \citet{1967ApOpt...6...51O} and \citet{2022MNRAS.511.4774O}.
We adopted a constant plate scale of 0.40{\arcsec} per pixel, which posses good signal-to-noise ratio 
in each spaxel. 
We evaluated the off-center value in RA and DEC relative to a reference position 
by measuring the center coordinate of the standard star in the 3-D datacube: 
the offset did not exceed 0.02{\arcsec} (0.05\,pixel) in both directions over the wavelengths. 

\begin{figure*}
\includegraphics[width=\textwidth]{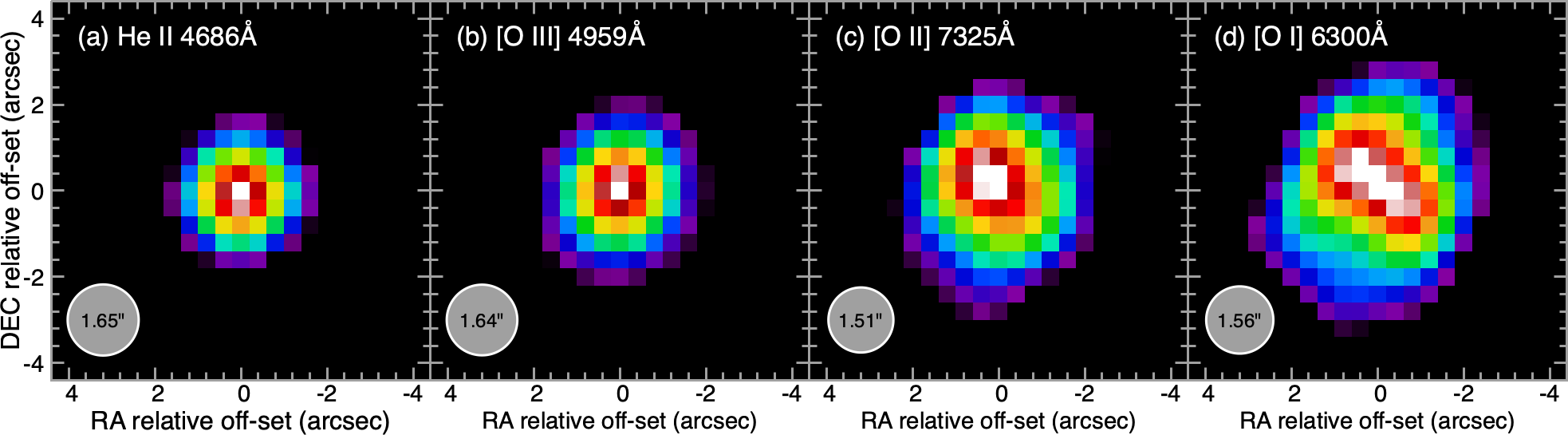}
\caption{Selected PSF-deconvolved KOOLS-IFU emission-line images. 
The intensity range is between the minimum and the maximum. 
The achieved spatial resolution is indicated by 
the gray circle. 
\label{F:map-kools}} 
\end{figure*}

\begin{table}
\caption{Summary of spectroscopic observations. \label{T-obslog}}
\centering
\begin{tabularx}{\columnwidth}{@{\extracolsep{\fill}}c@{\hspace{5pt}}c@{\hspace{3pt}}D{-}{-}{-1}c@{\hspace{3pt}}c}
\midrule	 
Date & Telescope/Inst. & \multicolumn{1}{c}{Range ({\micron})} &$R$\\
\midrule
2023/03/13 & Seimei/KOOLS-IFU  &0.420~-~0.853      &$\sim500$     \\
2023/04/10 & Seimei/KOOLS-IFU  &0.420~-~0.853        &$\sim500$     \\
2023/02/11 & Seimei/KOOLS-IFU  &0.560~-~1.010          &$\sim800$     \\
1983/08/03 & \emph{IUE}/SWP                     &0.115~-~0.198           &$\sim300$   \\
1983/08/03 & \emph{IUE}/LWR                      &0.185~-~0.335           &$\sim400$   \\
2007/06/17 & \emph{Spitzer}/IRS                 &5.2~-~14.5         &$\sim100$    \\
2007/06/17 & \emph{Spitzer}/IRS             &9.9~-~37.2          &$\sim600$   \\
\midrule
\end{tabularx}
\end{table}

For flux calibration and telluric removal, we synthesized the non-LTE theoretical spectrum of 
Hz44 using the {\sc Tlusty} code \citep{1988CoPhC..52..103H}. 
We adopted the basic stellar parameters from \citet{2019A&A...630A.130D} who determined 
surface gravity $\log_{10}\,g = 5.64$\,cm\,s$^{-2}$ and $T_{\rm eff} = 39100$\,K. We calculated the color excess between B and V bands $E(B-V)$ of 0.036 toward Hz44 in 
the total-to-selective extinction $R_{\rm V} = 3.1$,  using a comparison between the synthesized spectrum and the 
photometry spectral energy distribution (SED) in the range from 0.35 to 0.92\,{\micron} generated from 
\citet{2019A&A...621A..38G} and \citet{2010AJ....139.1242A}. In this procedure, we adopted 
the reddening law of \citet{1989ApJ...345..245C}. The flux calibration and telluric removal were performed using the function 
derived by comparing this reddened {\it telluric free} theoretical spectrum and the 1-D spectrum extracted from the observed 
3-D datacube. In the 1-D spectrum extraction, we recovered the lost flux outside the datacube 
using the Hz44's PSF in each wavelength. 

We evaluated the achieved spatial-resolution 
by measuring the Moffat function’s FWHM of the Hz44’s PSF every 
wavelength. The Moffat function is represented by the following: 

\begin{eqnarray}
I(r) &=& \left[1 + \left(r/{\alpha}\right)^{2}\right]^{-\beta} \\
{\alpha} &=& {\rm FWHM}\left[2^{\left(1/{\beta}\right)} - 1 \right]^{-0.5}, 
\label{E:moffat}
\end{eqnarray}

\noindent where $I(r)$ is the radial profile of these PSF standard stars, 
and $\beta$ is the shape parameter, set to be 4.0 through the wavelengths. 
To correct the image quality degradation due to airmass, we adopted the equation established 
by \citet{2022MNRAS.511.4774O}. Thus, the expected Moffat PSF's FWHM 
in arcsec can be expressed as a function of $\lambda$ in {\micron} using equation\,(\ref{E:psf}). 

\begin{eqnarray}
{\rm FWHM} = \left\{ \begin{array}{ll}
(1.70\pm0.01)\,(\lambda/0.55)^{-0.16\pm0.01} & \mbox{(Feb)}\\
(2.28\pm0.01)\,(\lambda/0.55)^{-0.20\pm0.01} & \mbox{(Mar)}\\
(1.88\pm0.01)\,(\lambda/0.55)^{-0.21\pm0.01} & \mbox{(Apr)}
\end{array} \right.
\label{E:psf}
\end{eqnarray}

After performing PSF-matching in all spectral bins in 
the datacube with equation\,(\ref{E:psf}), we carried out spectral analysis.

\begin{figure*}
\includegraphics[width=\textwidth]{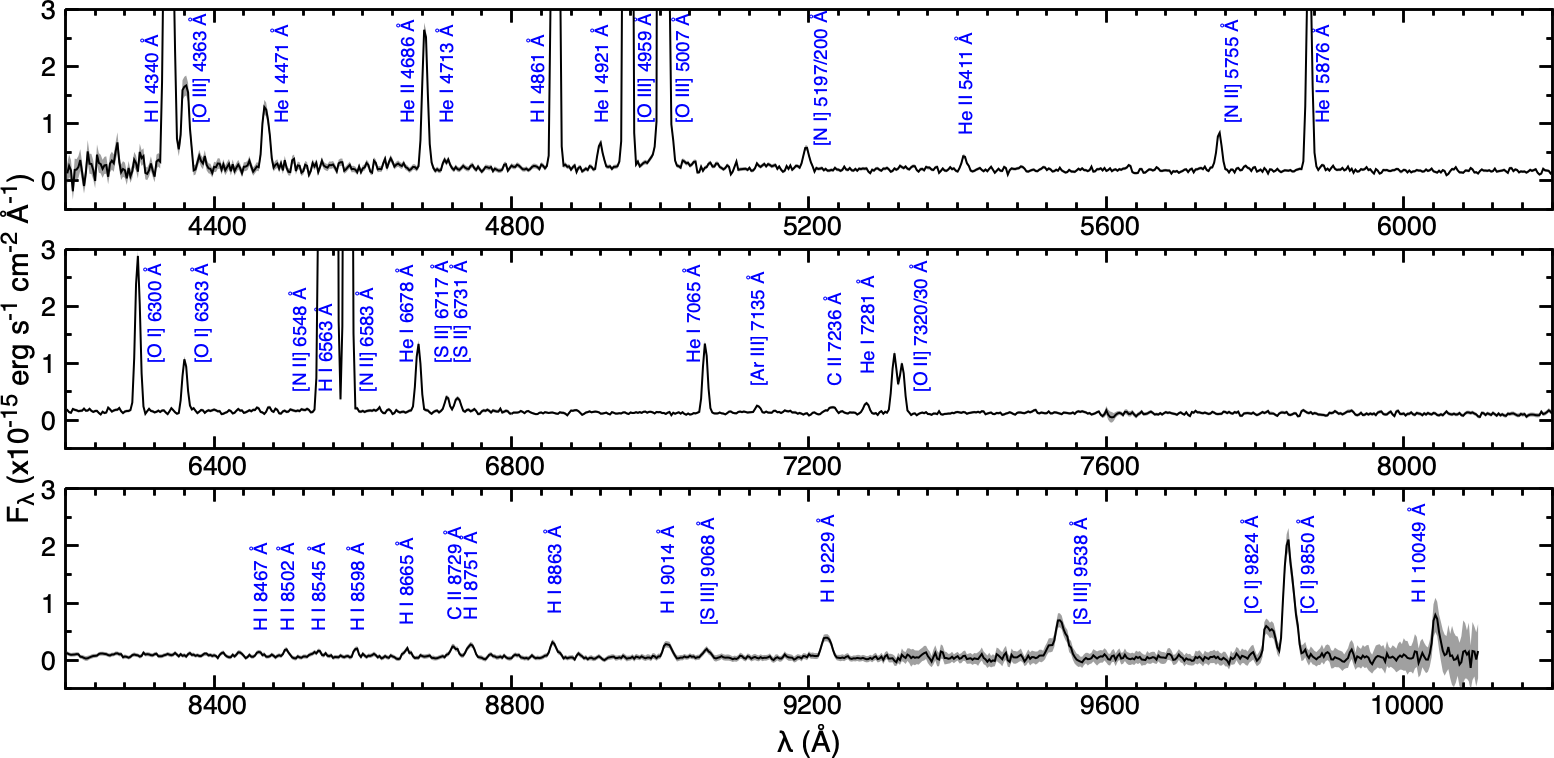}
\caption{The KOOLS-IFU 1-D spectrum. Prominent lines among 
the identified emission-lines are indicated in blue. The wavelengths are in air and also have a heliocentric correction. 
The dark-gray color shade indicates the one-sigma uncertainty at each wavelength. \label{F:spec-kools}} 
\end{figure*}

\subsection{\emph{IUE} UV and \emph{Spitzer} mid-IR spectra \label{S-irspec}}

We downloaded the SWP20599 and LWR16513 spectra (IUE Program ID: FA083, PI: M.~J.~Barlow) obtained by 
the International Ultraviolet Explorer (\emph{IUE})
from the Mikulski Archive for Space Telescopes (MAST). 
The size of the slit (9.9{\arcsec}~$\times$~22{\arcsec}) covers the entire nebula of H4-1. 
We combined these two spectra into a single $1150-3348$\,{\AA} spectrum. 

We used the mid-IR {\it Spitzer}/IRS \citep{2004ApJS..154...18H} 
short-low (SL), short-high (SH), and long-high (LH) spectra used 
in \citet{Otsuka:2013aa}. The size of the slit is 
$3.7\arcsec \times 57\arcsec$ in SL, $4.7\arcsec \times 11.3\arcsec$ in SH, and $11.1\arcsec \times 22.3\arcsec$ in LH. We combined these spectra into a single spectrum by scaling their flux densities to match with mid-IR  Wide-field Infrared Survey Explorer \citep[\emph{WISE};][]{2013wise.rept....1C} bands 3 and 4. More detailed information on the data and their reduction can be found therein.

\subsection{Multiwavelength Photometry}

We obtained broadband UV data from \emph{GALEX} \citep{2017ApJS..230...24B} and \emph{XMM-Newton} \citep{2012MNRAS.426..903P}, 
optical data from SDSS \citep{2015ApJS..219...12A}, 
near-IR data from UKIRT/UKIDSS \citep{2007MNRAS.379.1599L}, 
mid-IR data from \emph{Spitzer}/IRAC/MIPS
\citep[][for IRAC and MIPS, respectively]{2004ApJS..154...10F,2004ApJS..154...25R} and \emph{WISE}, and 
far-IR data from \emph{Herschel}/PACS \citep{2010A&A...518L...2P} as  
summarized in table\,\ref{T-obsphot}. The second column provides the extinction function $f(\lambda)$ at $\lambda$ 
computed using the reddening law of \citet{1989ApJ...345..245C} for the wavelengths 
shorter than UKIDSS 2.2\,{\micron} and \citet{1994A&AS..105..311F} for the wavelengths 
longer than IRAC 3.6\,{\micron} band. 
Herein, we adopt $R_{\rm V}$ of 3.1. The third and fourth columns in table\,\ref{T-obsphot}
are the dust extincted flux density $F_{\lambda}$ and the extinction-free 
one $I_{\lambda}$ obtained using equation\,(\ref{eq-1}) in section\,\ref{S-ext}, respectively.

\section{Results of the KOOLS-IFU observations}
\label{S-image}

First, we demonstrate the accuracy of the KOOLS-IFU datacube as photometric data. 
We measured the SDSS $r$-band $F_{\lambda}$ using the corresponding band 
image synthesized from the KOOLS-IFU data cube and the SDSS-$r$ band 
filter transmission. The resultant KOOLS-IFU SDSS $r$-band $F_{\lambda}$ 
is $(1.75\pm0.07)\times10^{-15}$\,{\ergs} while the cataloged SDSS $r$-band $F_{\lambda}$ 
is $1.76\times10^{-15}$\,{\ergs} (table\,\ref{T-obsphot}). 
This means our reduced KOOLS-IFU datacube has sufficient photometric accuracy.

In figure\,\ref{F:map-kools}, we present the PSF-deconvolved emission-line images. The signal-to-noise ratio per pixel is $\ge 5$. We generated these images by measuring the line fluxes in each spectrum every spaxel using the automated line fitting algorithm code \citep[{\sc Alfa};][]{Wesson:2016aa}. The achieved spatial resolution is 
evaluated by the Gaussian FWHM of the PSF deconvolved standard stars. 

The ionization stratification in H4-1 appears to be consistent with what is expected in a gaseous nebula surrounding the central ionization source;
that is, the spatial extent of gas line emission becomes larger
as the ionization energy of the emission becomes lesser.
The PSF-deconvolved Gaussian 1-sigma and the ellipticity of {\hb} emission are 0.87{\arcsec} and 0.06, respectively. They are 1.16{\arcsec} and 0.11 in the {\oi}\,6300\,{\AA}. 
The nebula is asymmetric and elongated along position angle of $\sim-25^{\circ}$. 
These KOOLS-IFU spectral images successfully resolved the spatially compact yet elongated emission core structure in the {\oii}\,7320/30\,{\AA} and {\oi}\,6300\,{\AA} lines. Such structure was resolved and confirmed only in near-IR H$_{2}$ images so far \citep{tajitsu:2023aa}. The core structure is one of the most distinctive spatial features seen in H4-1 and is considered to represent the flattened central equatorial disk in a bipolar PN. 

In figure\ref{F:spec-kools}, we show the spatially-integrated 1-D spectrum generated from 
the datacube. The flux density loss is corrected in every wavelength bin using the synthesized PSF of the 
standard star. The wavelength is in air and also has a heliocentric correction;
the heliocentric radial velocity measured using all the detected lines is 
$-181.8 \pm 4.9$\,{\kms}, which is consistent with \citet[$-181.35 \pm 0.35$\,{\kms}]{Otsuka:2013aa}. 
In the following sections, we use the KOOLS-IFU 1-D spectrum.

\section{Dust Extinction \label{S-ext}}

The observed flux $F$($\lambda$) (and flux density $F_{\lambda}$) is reddened by not only the 
interstellar and also the circumstellar dust. 
We obtain the extinction-free flux $I$($\lambda$) (and flux density $I_{\lambda}$) using equation\,(\ref{eq-1}):

\begin{eqnarray}
  I(\lambda) &=& F(\lambda)~\cdot~10^{c({\rm H\beta})(1 + f(\lambda))},
   \label{eq-1}
\end{eqnarray}

\noindent where $c$({\hb}) ($= \log_{10}$\,$I$({\hb})/$F$({\hb})) is 
the line-of-sight logarithmic reddening coefficient at {\hb}\,4861.33\,{\AA}.
Because $c$({\hb}) equals to 1.45\,$E(B-V)$ 
under the assumption of $R_{\rm V}$ being 3.1, 
one may simply adopt 
the $E(B-V)$ value measured from the Galaxy extinction map \citep[e.g.,][]{2011ApJ...737..103S}; the $E(B-V)$ value toward H4-1 is 0.0082, 
which is 0.012 in terms of $c$({\hb}). 
However, this $E(B-V)$ obtained by the Galactic extinction map is 
the average global interstellar medium (ISM) over a large solid angle around H4-1 because the angular resolution of the map is not high enough. 
Therefore, we directly calculate $c$({\hb}) by comparing observed {\hi} line 
fluxes and corresponding extinction-free {\hi} line-fluxes that are 
calculated by the {\hi} emissivity of \citet{Storey:1995aa} 
under the assumption of Case B. It should be noted that the estimated $c$({\hb}) value corresponds 
to the sum of the line-of-sight reddening coefficient at 4861.33\,{\AA} 
by the circumstellar and interstellar media.

To calculate $c$({\hb}) using the {\hi} lines, the de-reddened {\hi} line fluxes must be calculated for comparison. 
Obviously, the de-reddened {\hi} line fluxes are computed as functions of 
{\Ne} and {\te} \citep{Storey:1995aa}. 
However, {\Ne} and {\te} are actually the results of plasma diagnostics, which require extinction-corrected line fluxes. 
To make matters even worse, obtaining 
the accurate $c$({\hb}) value in highly excited nebulae showing {\heii} lines, 
we need to subtract the contaminating neighboring {\heii} line contribution to {\hi} lines.

This ``Chicken or the Egg'' dilemma \citep[for details, see][]{2021PASP..133i3002U,2022Galax..10...30U} and {\heii} line contamination in $c$({\hb}) calculation can be simultaneously 
resolved via an iterative loop by equations\,(\ref{eq-2}) -- (\ref{eq-8}).
The He\,{\sc ii} line subtraction processes are included in equations\,(\ref{eq-2}) -- (\ref{eq-4}). 

\begin{eqnarray}
F({{\rm H}\,{\rm I}}_{\lambda_i,j}) &=& F({{\rm H}\,{\rm I}}_{\lambda_{i,j}})_{\rm obs} - F({\rm He}\,{\rm II}\,4686)_{\rm obs} \nonumber\\
           &&                        \cdot(I({{\rm He}\,{\rm II}}_{\lambda_{i,j}})/I({\rm He}\,{\rm II}\,4686))_{(n_{\rm e}, T_{\rm e})}  \nonumber\\
           &&                        \cdot10^{c({\rm H}\beta)_{i,j} ( f(4686) - f({\rm He}\,{\rm II}_{\lambda_{i,j}})) } \label{eq-2}\\
F({\rm H}\alpha) &=& F({\rm H}\alpha)_{\rm obs} - F({\rm He}\,{\rm II}\,4686)_{\rm obs} \nonumber\\
           &&                        \cdot(I({{\rm He}\,{\rm II}}\,6560)/I({\rm He}\,{\rm II}\,4686))_{(n_{\rm e}, T_{\rm e})}  \nonumber\\
           &&                        \cdot10^{c({\rm H}\beta)_{j} ( f(4686) - f({\rm He}\,{\rm II}\,6560)) } \label{eq-3}\\
F({\rm H}\beta) &=& F({\rm H}\beta)_{\rm obs} - F({\rm He}\,{\rm II}\,4686)_{\rm obs} \nonumber\\
           &&                          \cdot(I({{\rm He}\,{\rm II}}\,4860)/I({\rm He}\,{\rm II}\,4686))_{(n_{\rm e}, T_{\rm e})}  \nonumber\\
           &&                        \cdot10^{c({\rm H}\beta)_{i} ( f(4686) - f({\rm He}\,{\rm II}\,4860)) } \label{eq-4}\\
c({\rm H}\beta)_{i} &=& \frac{1}{f(\lambda_i)}\log_{10}\frac{(I({{\rm H}\,{\rm I}}_{\lambda_i})/I({\rm H}\beta))_{(n_{\rm e}, T_{\rm e})}}{F({{\rm H}\,{\rm I}}_{\lambda_i})/F({\rm H}\beta)} \\
c({\rm H}\beta)_{j} &=& \frac{1}{f(\lambda_j)}\log_{10}\frac{(I({{\rm H}\,{\rm I}}_{\lambda_j})/I({\rm H}\alpha))_{(n_{\rm e}, T_{\rm e})}}{F({{\rm H}\,{\rm I}}_{\lambda_j)}/F({\rm H}\alpha)} \\
c({\rm H}\beta) &=& \frac{\sum_{i,j} c({\rm H}\beta)_{i,j}{\sigma}_{i,j}^{-2}}{\sum_{i,j}{\sigma}_{i,j}^{-2}} \\
\delta c({\rm H}\beta)^{2} &=& \sum_{i,j}{\sigma}_{i,j}^{-2}, \label{eq-8}
\end{eqnarray}

\noindent where $F({{\rm H}\,{\rm I}}_{\lambda_i,j})_{\rm obs}$ is the observed line-flux of the {\hi} and {\heii} at $\lambda_{i,j}$, 
$F({{\rm H}\,{\rm I}}_{\lambda_i,j})$ is the sole {\hi} line-flux at $\lambda_{i,j}$, 
$I$(HI$_{\lambda_{i}}$)/$I$({\hb})$_{(n_{\rm e}, T_{\rm e})}$ and $F$(HI$_{\lambda_{i}}$)/$F$({\hb}) are the 
{\heii} contamination-free theoretical and observed ratios of {\hi} line at $\lambda_{i}$ with respect to {\hb} 
in {\Ne} and {\te}, and $I$(HI$_{\lambda_{j}}$)/$I$({\ha})$_{(n_{\rm e}, T_{\rm e})}$ and $F$(HI$_{\lambda_{j}}$)/$F$({\ha}) are the theoretical 
and observed ratios of {\hi} line at $\lambda_{j}$ wrt. {\ha} in {\Ne} and {\te}.
$I$({\heii}$_{\lambda_{i,j}}$)/$I$({\heii}\,4686) is the theoretical ratio of {\heii} at $\lambda_{i,j}$ wrt. {\heii}\,4686\,{\AA}. 
$\sigma_{i,j}$ is the standard deviation of $c$({\hb})$_{i,j}$. 
We set the searching range of {\te} and {\Ne} to be $5000-15000$\,K and 
$100-5000$\,cm$^{-3}$. 

In practice, due to the lack of the available higher order {\hi} lines which 
are sensitive to $n_{\rm e}$, we used {\sii}\,6717/31\,{\AA} line ratio as the density constraint. 
Adopting a genetic algorithm, we obtained the {\te}--{\Ne} pair whose values 
give the minimum of $\delta$\,$c$({\hb})$^{2}$ after 100 times independent calculations.
Finally, we obtained $c$({\hb}) = $0.103 \pm 0.024$ 
in {\te} = $12635 \pm 1412$\,K and {\Ne} = $420 \pm 38$\,cm$^{-3}$.

\section{Chemical Abundances \label{S:abund}}

\begin{table}
\caption{Derived {\Ne}, {\te}, and ionic/elemental abundances. \label{T-abund}}
\centering
\begin{tabularx}{\columnwidth}{@{\extracolsep{\fill}}l@{\hspace{0pt}}D{p}{\pm}{-1}}
\midrule	 
Parameter &\multicolumn{1}{c}{Value}\\
\midrule	 
{\te}({\oiii}) &12885~p~72\,({\rm K})\\
{\te}({\nii})         &12976~p~46\,({\rm K})\\
{\te}({\hei})          &7456~p~1774\,({\rm K})\\
{\Ne}({\siii})  &3074~p~2311\,({\rm cm}^{-3})\\
{\Ne}({\sii})                &441~p~158\,({\rm cm}^{-3})\\
\midrule	 
Ion X$^{i+}$ ($\#$ of the used lines) &\multicolumn{1}{c}{$n$(X$^{i+}$)/$n$(H$^{+}$)}\\
\midrule	 
He$^{+}$ (5)& 9.50\times10^{-2}~p~3.53\times10^{-3} \\
He$^{2+}$ (3) &8.81\times10^{-3}~p~1.74\times10^{-4} \\ 
C$^{0}$(CEL) (2)& 2.64\times10^{-6}~p~1.54\times10^{-7} \\ 
C$^{+}$(CEL) (1)& 1.23\times10^{-4}~p~1.13\times10^{-5} \\ 
C$^{2+}$(CEL) (1)& 5.31\times10^{-4}~p~4.23\times10^{-5} \\ 
C$^{2+}$(RL) (1)& 4.07\times10^{-4}~p~8.71\times10^{-5} \\ 
C$^{3+}$(CEL) (1)& 7.63\times10^{-5}~p~7.04\times10^{-6} \\ 
N$^{0}$  (1)& 4.81\times10^{-7}~p~4.19\times10^{-8} \\ 
N$^{+}$ (3)& 1.13\times10^{-5}~p~2.29\times10^{-7} \\ 
O$^{0}$ (1)& 8.78\times10^{-6}~p~2.85\times10^{-7} \\ 
O$^{+}$ (2)& 3.07\times10^{-5}~p~2.84\times10^{-6} \\ 
O$^{2+}$ (3)& 1.01\times10^{-4}~p~1.67\times10^{-6} \\ 
O$^{3+}$ (1)& 7.34\times10^{-6}~p~3.52\times10^{-6} \\ 
Ne$^{+}$ (1)& 4.07\times10^{-7}~p~5.11\times10^{-8} \\ 
Ne$^{2+}$ (1)& 1.91\times10^{-6}~p~1.29\times10^{-7} \\ 
S$^{+}$ (2)& 2.69\times10^{-8}~p~1.36\times10^{-9} \\ 
S$^{2+}$ (3)& 8.25\times10^{-8}~p~6.96\times10^{-9} \\ 
S$^{3+}$ (1)& 2.99\times10^{-8}~p~4.71\times10^{-9} \\ 
Ar$^{2+}$ (1)& 2.42\times10^{-8}~p~2.81\times10^{-9} \\ 
\midrule	 
Element X        &\multicolumn{1}{c}{$n$(X)/$n$(H)} \\
\midrule	 
He &1.04\times10^{-1}~p~3.53\times10^{-3}\\
C(CEL)  &7.31\times10^{-4}~p~4.44\times10^{-5}\\
N  &5.45\times10^{-5}~p~5.46\times10^{-6}\\
O  &1.47\times10^{-4}~p~4.83\times10^{-6}\\
Ne &2.60\times10^{-6}~p~1.89\times10^{-7}\\
S  &1.39\times10^{-7}~p~8.51\times10^{-9}\\
Ar &4.09\times10^{-8}~p~6.38\times10^{-9}\\
\midrule	 
Element X        &\multicolumn{1}{c}{12 + $\log_{10}$\,$n$({\rm X})/$n$({\rm H})} \\
\midrule	 
He &11.02~p~0.01\\
C(CEL)  &8.86~p~0.03\\
N  &7.74~p~0.05\\
O  &8.17~p~0.01\\
Ne &6.42~p~0.04\\
S  &5.14~p~0.03\\
Ar &4.61~p~0.07\\
\midrule
\end{tabularx}
\end{table}

We measured the line-fluxes of atomic species appeared in the 1-D KOOLS-IFU, \emph{IUE}, and \emph{Spitzer} spectra using 
{\sc Alfa} \citep{Wesson:2016aa}. 
For several lines, we manually measured their fluxes by fitting a multiple Gaussian function to their line-profiles. 
In table\,\ref{T-line}, we summarize 
the resultant dereddened line-fluxes normalized to the line-flux of {\hb} whose intensity is 100. 
Here, the dereddened {\hb} line-flux $I$({\hb}) is ($4.54 \pm 0.25$)$\times10^{-13}$\,erg\,s$^{-1}$\,cm$^{-2}$.

To calculate elemental abundances, {\te}, {\Ne}, and then ionic abundance derivations are required. 
The {\Ne} and {\te} values using collisionally excited lines (CELs) are determined at the intersection 
of the {\Ne} and {\te} diagnostic curves in each ionization zone presented in table\,\ref{T-abund}. 
In the neutral/low-ionization zone where the neutral or singly ionized lines would be 
emitted, {\te}({\nii}) and {\Ne}({\sii}) are determined from their curves. 

In the high-ionization zone where 
the doubly or more ionized lines would be emitted, {\te}({\oiii}) and {\Ne}({\siii}) are derived from their curves. {\te}({\nii} and {\Ne}({\sii}) are derived from the 
{\nii} ($I$(6548\,{\AA})+$I$(6583\,{\AA}))/$I$(5755\,{\AA}) and 
{\sii} $I$(6717\,{\AA})/$I$(6731\,{\AA}). 
 {\te}({\oiii} and {\Ne}({\siii}) are derived from the 
{\oiii} ($I$(4959\,{\AA})+$I$(5007\,{\AA}))/$I$(4363\,{\AA}) and 
{\siii} $I$(18\,{\micron})/$I$(9068\,{\AA}). Because of no detection of the $[$N\,{\sc iii}$]$\,1744-54\,{\AA} lines in the \emph{IUE} spectrum, we did not consider 
the recombination contribution of the N$^{2+}$ to the {\nii}\,5755\,{\AA} line. 
The contributions of the O$^{2+}$ and O$^{3+}$ to the {\oii}\,7320/30\,{\AA} and {\oiii}\,4363\,{\AA} lines are not subtracted because their contributions are very small as reported by \citet{Otsuka:2013aa}. 
{\te}({\hei}) is derived from the {\hei} $I$(7281\,{\AA})/$I$(6678\,{\AA}) ratio and 
the {\te}--{\Ne} logarithm interpolate function of the {\hei} effective recombination coefficients by \cite{2022MNRAS.511.4774O}. In the calculation, we adopt {\Ne}({\sii}) and also subtract the {\heii}\,6680\,{\AA} line-flux from the complex of the {\hei}\,6678\,{\AA} plus {\heii}\,6680\,{\AA} lines based on the theoretical ratio of the 
{\heii} $I$(6680\,{\AA})/$I$(4686\,{\AA}) under {\te}({\oiii}) and {\Ne}({\siii}).

The ionic abundance $n$(X$^{i+}$)/$n$(H$^{+}$) and elemental abundances are summarized in table\,\ref{T-abund}. In the CEL ionic abundances,
the values are obtained by solving the equations of the populations at multiple energy levels under the obtained CEL {\te} and {\Ne}. When two or more lines are available, we calculate the ionic abundance using each line. Then, we perform the weight-average among their values.
In calculations of X$^{i+}$/H$^{+}$, {\te} and {\Ne} of H$^{+}$ and co-existing X$^{i+}$ are the same.
$n$(He$^{+}$)/$n$(H$^{+}$) is calculated using the {\te}({\hei}) and {\Ne}({\sii}) values.  
$n$(He$^{2+}$)/$n$(H$^{+}$) is calculated using the {\te}({\oiii}) and {\Ne}({\siii}) values. 
$n$(C$^{2+}$)/$n$(H$^{+}$) from the recombination line {\cii}\,7236\,{\AA} is calculated using the {\te}({\oiii}) and {\Ne}({\siii}) values.
In the calculation of the CEL ionic abundances, we adopt {\te}({\nii}) and {\Ne}({\sii}) for the neutral or singly ionized species, and {\te}({\oiii}) and {\Ne}({\siii}) for the doubly or more ionized ones.

Finally, we calculate the elemental abundances $n$(X)/$n$(H) by introducing the ionization correction 
factor ICF(X) to element X. ICFs recover the ionic abundances in unobserved ionization 
stages covered by the obtained spectra. The elemental abundance is calculated by  
ICF(X)~$\cdot$~$\sum$ $n$(X$^{i+}$)/$n$(H$^{+}$). We determine ICF(X) by referring to \citet{Otsuka:2013aa}. 
For He, C, O, and S, their elemental abundances are derived by simply summing up the derived ionic abundances. 
ICF(N) is equal to the O/O$^{+}$ value ($4.82 \pm 0.50$). ICF(Ne) corresponds to the O/(O$^{+}$ + O$^{2+}$) value 
($1.13 \pm 0.05$). 
ICF(Ar) is the S/S$^{2+}$ value ($1.70 \pm 0.19$). 

H4-1 is proven to be an extremely C-rich and metal-poor PN; 
the respective CEL [C,N,O,Ne,S,Ar/H] are $+0.42 \pm 0.05$, $-0.09 \pm 0.07$, $-0.52 \pm 0.05$, $-1.50 \pm 0.11$, $-1.97 \pm 0.04$, and 
$-1.78 \pm 0.15$, where \cite{Asplund:2009aa} is adopted as the 
solar abundance. 
The CEL C abundance (12 + $\log_{10}n({\rm C})/n({\rm H})$) shows large scatter among previous works, 
from  8.68\, dex in \cite{2003PASP..115...80K} to 9.02\, dex in \citet{Otsuka:2013aa}. 
This variation can be attributed to the difference in the slit-entrance size adopted between the \emph{IUE} observations and the ground-based long-slit observations, and the accuracy of aperture correction is compromised. As our KOOLS-IFU study took the aperture correction into account, our derived CEL C abundance  is likely to be more reliable than previous estimates. 
We noticed a large difference in the N abundance 
(12 + $\log_{10}n({\rm N})/n({\rm H})$) in comparison to 7.56\, dex, previously reported by \citet{Otsuka:2013aa}. This is probably because the slit-length used in their Subaru/HDS observations was not long enough to cover the neutral/singly ionized regions where {\nii} lines are emitted.

\section{Gas-to-Dust mass ratio, gas and dust masses}
\label{S-gdr}

\begin{table*}
\caption{Comparison of elemental abundances between the 
models by \citet{2012ApJ...747....2L} and our observation. 
The initial [$\alpha$/Fe] in the 
2.0\,M$_{\sun}$ model are the same as the prediction by the Galaxy chemical evolution model by \citet{2011MNRAS.414.3231K}.
\label{T-abund2}}
\begin{tabularx}{\textwidth}{@{\extracolsep{\fill}}l@{\hspace{3pt}}
D{.}{.}{-1}@{\hspace{3pt}}
D{.}{.}{-1}@{\hspace{3pt}}
D{.}{.}{-1}@{\hspace{3pt}}
D{.}{.}{-1}@{\hspace{3pt}}
D{.}{.}{-1}@{\hspace{3pt}}
D{.}{.}{-1}}
\midrule
Model of initial mass&\multicolumn{1}{c}{C}   &\multicolumn{1}{c}{N}   &\multicolumn{1}{c}{O}    
&\multicolumn{1}{c}{Ne}  &\multicolumn{1}{c}{S}   &\multicolumn{1}{c}{Ar}\\
\midrule                      
0.89\,M$_{\sun}$ (w/o PMZ)&9.07&7.59&7.48 &7.33&5.00&4.28\\
1.00\,M$_{\sun}$ (w/o PMZ)&8.15&6.56&6.97 &7.33&5.00&4.26\\
1.25\,M$_{\sun}$  (w/o PMZ) &8.94&6.66&7.31 &6.95&5.00&4.27\\
1.25\,M$_{\sun}$  (w/\phantom{o} PMZ)&8.90&6.68&7.47 &7.65&5.00&4.27\\
1.50\,M$_{\sun}$   (w/o PMZ)  &9.26&6.80&7.56 &7.68&5.01&4.29\\
2.00\,M$_{\sun}$  (w/\phantom{o} PMZ)&9.55&6.84&7.93 &8.66&5.34&4.57\\
\midrule
H4-1             &8.86&7.74&8.17 &6.42&5.14&4.61\\
\midrule
\end{tabularx}
\end{table*}

The $c$({\hb}) value is the sum of the line-of-sight ISM and circumstellar logarithmic reddening coefficient at 4861.33\,{\AA}. 
By subtracting the ISM $c$({\hb}) value (0.01) from the derived 
$c$({\hb}) value (section\,\ref{S-ext}), we obtain the reddening coefficient due only to circumstellar media. Thus, we derive the circumstellar gas-to-dust mass ratio (GDR). 
\emph{Spitzer}/IRS spectrum exhibits (1) emission bands at $5-9$\,{\micron}, 11.3\,{\micron}, and 12.7\,{\micron} which 
would be attributed to poly-aromatic amorphous carbon (PAH) molecules and 
(2) feature-less dust continuum which would be attributed to amorphous carbon grains. 
Here, we estimate the GDR $\psi$ and the gas/dust masses $m_{\rm g}$/$m_{\rm d}$ 
using equations\,(7)--(13) established by \citet{2022MNRAS.511.4774O}. 

When gas and dust grains coexist in the same volume, $\psi$ can be expressed by equation\,(\ref{eq-9}):

\begin{eqnarray}
\psi &=& \frac{ 3 (\log\,e) Q_{\rm ext}({\rm H}\beta,a) C}
{ 4 a \rho_{\rm d} c({\rm H}\beta)_{\rm cir.}}
\sum 
\frac{4 \pi I_{g} \mu_{g} }{n_{\rm e} j_{\rm g}(T_{\rm e},n_{\rm e}) },
\label{eq-9}
\end{eqnarray}

\noindent where $Q_{\rm ext}$({\hb},$a$) is the extinction cross section of the grain of radius $a$ at {\hb}, and
$\rho_{\rm d}$ and $a$ are the density and the radius of the grain, respectively.
$C$ is a dimensionless constant: when assuming the spherical shaped nebula 
with the radius of $r$ in arcsec, $C$ equals to $\pi^{-1}(4.848\times10^{-6}~r)^{-2}$ \citep{2022MNRAS.511.4774O}. 
$c$({\hb})$_{\rm cir.}$ is the line-of-sight reddening coefficient due to circumstellar media, corresponding 
to $c$({\hb})$_{\rm tot.}$ (the sum of the line-of-sight ISM and circumstellar 
reddening coefficient, i.e., $0.103 \pm 0.024$) minus $c$({\hb})$_{\rm ISM}$ (due to ISM, i.e., 0.01). 
$n_{\rm g}$ and $\mu_{\rm g}$ are the number density and atomic weight of the target gas, respectively. 
$j_{\rm g}$({\te},{\Ne}) is the volume emissivity of the gas component per $n_{\rm g}n_{\rm e}$.
$I_{\rm g}$ is the observed extinction-free line flux of the gas component that is corrected using equation\,(\ref{eq-1}) with 
$c$({\hb})$_{\rm tot.}$.

In calculating GDR, we utilize the {\hi}, {\hei}, and {\heii} lines.
{\te} = $12634 \pm 1412$\,K and {\Ne} = $420 \pm 38$\,cm$^{-3}$ are adopted for the H$^{+}$ mass, 
{\te}({\hei}) and {\Ne}({\sii}) for the He$^{+}$ mass,
{\te}({\oiii}) and {\Ne}({\siii}) for the He$^{2+}$ mass. 
We adopt $r = 1.5${\arcsec} estimated from the H$_{2}$ radial profile presented in \citet{tajitsu:2023aa}.
We also adopt an AC-type (a-C:H) amorphous carbon grain 
($\rho_{\rm d}$ of 1.85\, g\,cm$^{-3}$) with the radius of $0.01-0.50$\,{\micron} and $a^{-3.5}$ size distribution 
(the average $a$ of 0.07\,{\micron}). 
$\langle Q_{\rm ext}({\rm H}\beta,0.01-0.50\,{\micron}) \rangle$ of 0.88 from the optical data of \cite{Rouleau:1991aa}. 
Then, we obtain GDR of $387 \pm 86$ including the uncertainty of line flux, 
c({\hb}), {\te}, and {\Ne}.

Next, we directly calculate the total atomic gas mass $m_{\rm g}$ 
of $(1.28 \pm 0.24)\times10^{-3}$\,$D_{\rm kpc}^{2}$\,M$_{\sun}$ at 
the distance of $D_{\rm kpc}$ in kpc, by simply adding the H$^{+}$ and He$^{+,2+}$ gas masses derived from equation\,(\ref{eq-10}):

\begin{eqnarray}
m_{\rm g} &=& \sum \frac{4 \pi D^{2} I_{g} \mu_{g} }{n_{\rm e} j_{\rm g}(T_{\rm e},n_{\rm e}) }.
\label{eq-10}
\end{eqnarray}

The respective H and C gas masses are $(9.04 \pm 1.70)\times10^{-4}$\,$D_{\rm kpc}^{2}$\,M$_{\sun}$ and 
$(7.93 \pm 1.57)\times10^{-6}$\,$D_{\rm kpc}^{2}$\,M$_{\sun}$. 
Adopting the determined $m_{\rm g}$, we obtain the dust mass $m_{\rm d}$ of 
$(3.31 \pm 0.62)\times10^{-6}$\,$D_{\rm kpc}^{2}$\,M$_{\sun}$. 
Note that the derived $m_{\rm g}$ and $m_{\rm d}$  here indicate the fully ionized gas mass and the dust mass co-existing with it. 
The total mass of the C atom in the gas phase and that incorporated into the dust grains is estimated to be $(1.12 \pm 0.17)\times10^{-5}$\,$D_{\rm kpc}^{2}$\,M$_{\sun}$. This means $\sim$30\,$\%$ of the total C atom synthesized by the progenitor 
could exist as dust grains. As a reference, 12 + $\log_{10}$\,$n$(C)/$n$(H) would be $9.02 \pm 0.11$ when including the C atom 
locked in the dust grains. Through the photoionization model using 
the derived GDR value ($387 \pm 86$) as a constraint, we will estimate the gas mass including the neutral hydrogen and molecules later.

\section{Discussions}
\label{S-dis}

\subsection{The possible origin and evolution of H4-1\label{S-dis-ori}}

The initial progenitor mass range can be determined by comparing the observed elemental abundances 
with their predicted counterparts taken from AGB nucleosynthesis models for a single star
(\citealt{2012ApJ...747....2L}; table\,\ref{T-abund2}). 
Note that the values in H4-1 are in the gas phase. 
As the available AGB nucleosynthesis models for stars with $Z = 10^{-4}$ are only those by \citet{2012ApJ...747....2L}, we compare their model results with the observed elemental abundances.

Compared with the other initial mass models, the 0.89\,M$_{\sun}$ initial-mass model without a partial mixing zone (PMZ\footnote{%
In low-mass AGB stars, a PMZ between the bottom of the H-rich convective envelope and the outermost region of the $^{12}$C-rich intershell layer leads to a synthesis of extra $^{13}$C and $^{14}$N. 
Hence, the presence of a PMZ would yield more C, O, Ne, and neutron-capture elements than cases without a PMZ \citep[e.g.,][]{2010MNRAS.403.1413K}
}) is very close to the observed abundances.
However, the discrepancy in the Ne and Xe\footnote{
For example, the 0.89\,M$_{\sun}$ initial-mass model predicts 12+$\log_{10}$(Xe/H) = 1.54. \citet{Otsuka:2013aa} conclude that 
the observed Xe abundance can be explained by models with initially 
$r$-process element rich.} abundances is rather large even for this preferred single-star interpretation, as seen in table\,\ref{T-abund2} and previously determined \citep[$12+\log_{10}$(Xe/H) $> 2.75$;][]{Otsuka:2013aa}. We should also note that the evolutionary time scale 
since the end of the AGB phase 
for the 0.89\,M$_{\sun}$ initial-mass model appears much too long (table\,\ref{T-dis}). 
Thus, in practice, we need not consider the AGB nucleosynthesis model for stars of initially 0.89\,M$_{\sun}$.

The 1.00\,M$_{\sun}$ initial-mass model without PMZ 
predicts much lower C and O abundances (8.16 and 6.94\,dex, respectively) 
compared to the observed values. 
The 1.25\,M$_{\sun}$ initial-mass model without PMZ shows 
good agreement in terms of C and Ne abundances, 
whereas O and Xe (0.45\,dex) abundances are very low
(table\,\ref{T-abund2}, w/o PMZ).
On the other hand, the 1.25\,M$_{\sun}$ initial-mass model with PMZ 
(of $2.0\times10^{-3}$\,M$_{\sun}$; table\,\ref{T-abund2}, w/ PMZ) 
predicts more C, O, Ne, and Xe (1.45\,dex).
The 1.50\,M$_{\sun}$ initial-mass model without PMZ appears similar 
to the 0.89\,M$_{\sun}$ initial-mass model except for N and O. 
Previously, 
it was found that the observed elemental abundances agreed 
model predictions from the 2.00\,M$_{\sun}$ initial-mass model 
with PMZ of $2.0\times10^{-3}$\,M$_{\sun}$ and initially 
$\alpha$-element/$r$-process-element rich ([$\alpha$/Fe]$\sim+0.3$ and [$r$/Fe]$\sim+0.4$), 
except for C, N and Ne abundances
\citep{Otsuka:2013aa}. 
Based on the reduced-$\chi^{2}$ evaluation between the observed and predicted
C, N, O, Ne, and Xe abundances, 
the 1.25\,M$_{\sun}$ initial-mass star model without PMZ corresponds the most favorably to the observed abundances. Even if adopting the C abundance including the C atom locked in the dust grains ($9.02 \pm 0.11$\,dex), this initial mass model keeps good fit to the observed abundances. 
Hence, the progenitor mass of H4-1 could have been greater than $0.8-0.9$\,M$_{\sun}$,
but not so much more than $\sim2$\,M$_{\sun}$.

However, all single star models are still largely discrepant in terms of the N, O, and Ne abundances.
The high N abundance in H4-1 could be caused by extra-mixing during the red giant branch phase and also by helium-flash-driven deep mixing during the AGB phase. 
The latter process would produce $^{14}$N through the $^{13}$C($p$,$\gamma$)$^{14}$N reaction by mixing protons into the He-burning shell. 
Incorporating extra overshooting, which was not considered by \citet{2012ApJ...747....2L}, 
 more C and O could be dredged up from the core to the stellar surface (M.\,Lugaro, {\it priv.\ comm.}). 
However, this mechanism would also cause an Ne overabundance. 
To lower the Ne abundance, 
the progenitor star could start as Ne poor or extra mixing would have to be invoked ($^{22}$Ne abundance would be reduced; M.\,Lugaro, {\it priv.\ comm}). 
\citet{1989MNRAS.241..453H} interpreted that H4-1 formed in the low Ne environment.
We believed that the low Ne abundance can be improved by adding the Ne$^{+}$ abundance 
re-calculated using the PSF-matched KOOLS-IFU and {\it Spitzer} spectra. 
However, the Ne abundance remains low and also cannot be reproduced by 
single star AGB nucleosynthesis models. 
The extremely low Ne abundance in H4-1 seems to be a serious problem in the stellar nucleosynthesis and the Galaxy chemical evolution \citep{2004AJ....127.2284H}. 
Because there does not seem to be any single star models that simultaneously satisfy all of the observed N, O, and Ne abundances, other possibilities, e.g., binary star evolution, would have to be considered.

\citet{Otsuka:2013aa} already speculated that a binary system could explain the rather anomalous elemental abundances seen from H4-1. 
The apparently anomalous N, O, and Ne abundances of H4-1 could originate from its binarity.
According to \citet{2016A&A...588A..25M}, a single star of $0.85-2.5$\,M$_{\sun}$ initial mass is predicted to eventually evolve into a white dwarf of $\sim 0.53-0.75$\,M$_{\sun}$. 
This naively means that a binary system composed of a 2.5\,M$_{\sun}$ primary and a 0.85\,M$_{\sun}$ companion
would become a star of about 1.6\,M$_{\sun}$ via a merger event that takes place after the primary experienced mass loss.
Such a {\sl merged} system could exhibit abundances that are largely consistent with originally $0.8-2.5$\,M$_{\odot}$ single stars (table\,\ref{T-abund2}).

To further determine the origin of H4-1, we quantify
the total circumstellar mass based on the observational data. 
To that end, 
we first estimate the distance to H4-1 by assuming 
(i) a single 0.85\,M$_{\sun}$ initial-mass star and 
(ii) a binary system in which a 0.85\,M$_{\sun}$ main-sequence secondary 
merging with a white dwarf primary that has become a single main-sequence star of 1.25\,M$_{\sun}$ or a 1.5\,M$_{\sun}$.

\subsection{Distance toward H4-1}
\label{S-dist}

The distance to H4-1 has not yet been well constrained, 
from 4.4 to 25.3\,kpc with the average of 15.68\,kpc (1-sigma = 6.1\,kpc) in the literature 
\citep{1971ApJS...22..319C,1978A&AS...33..367A,1981A&A....94..365K,1984A&AS...55..253M,
1992A&AS...94..399C,1992MNRAS.257..317K,1995ApJS...98..659Z,1997ARep...41..760M,
2008ApJ...689..194S,2016MNRAS.455.1459F}. 
Assuming that the nebula is optically thick and absorbs all the energy emitted from the central star and 
re-emits it, the integrated flux from the central star corresponds to the sum of all the observed fluxes 
of the nebular atomic and molecular gas and dust grains and is proportional to the square of the distance. 
Therefore, even if no direct spectrum and photometry of the central star itself is available, 
the distance can still be estimated by comparing all the nebular-line/continuum flux of a target PN 
with those of a reference PN whose distance is well known.

We note, however, that this distance estimate is valid only 
when the target and reference PNe have similar initial mass/metal abundances 
and their central stars are in the same evolutionary stage.
As far as we are aware, 
K648 in M15 is the only candidate as a reference PN for H4-1:
both are halo PNe and 
the distance to K648 is well calibrated \citep{2015ApJS..217...22O}. 
However, 
K648 has a cooler central star ($T_{\rm eff} = 36360$\,K), 
still evolving toward the PN turn-off, 
while H4-1 has a hotter central star \citep[e.g., $T_{\rm eff} = 66000-99000$\,K;][]{2003PASP..115...67O} 
already in the white dwarf cooling track. 
Therefore, the distance to H4-1 obtained through this flux comparison method remains uncertain at best.

\begin{table}
\caption{The estimated distance ($D$) toward H4-1 for different initial masses, 
effective temperatures ($T_{\rm eff}$), luminosities ($L_{\ast}$), surface gravities ($\log_{10}\,g$) of the central star. We define the age as the time since when $T_{\rm eff}=5000$\,K. Dagger ($\dagger$; the first row for each initial mass star) indicates that the central star is still becoming hot, while the rest are later epochs on the cooling track. 
\label{T-dis}}
\centering
\begin{tabularx}{\columnwidth}{@{\extracolsep{\fill}}D{.}{.}{-1}@{\hspace{2pt}}c@{\hspace{5pt}}D{.}{.}{-1}@{\hspace{-10pt}}c@{\hspace{0pt}}D{.}{.}{-1}@{\hspace{2pt}}c}
\midrule
\multicolumn{1}{c}{Init. mass}&$T_{\rm eff}$ &\multicolumn{1}{c}{$L_{\ast}$~~~} &$\log_{10}\,g$ 
&\multicolumn{1}{c}{$D$} &Age\\
\multicolumn{1}{c}{(M$_{\sun}$)}   &(K)&\multicolumn{1}{c}{($L_{\sun}$)~~~}&(cm\,s$^{-2}$)&\multicolumn{1}{c}
{(kpc)} &($\times10^{4}$ yrs)\\
\midrule
0.85^{\dagger} &125000&1179.1 &6.43   &25.03&3.66\\
0.85 &125000&406.7  &6.89&14.70 &5.72\\
0.85 &122500&332.1  &6.96&13.28 &6.23\\
0.85 &120000&274.3  &7.00&12.07 &6.59\\
\midrule
1.25^{\dagger} &125000&7139.9 &5.70&62.21&0.34\\
1.25 &125000&174.4  &7.31&9.63&0.49\\
1.25 &122500&156.8  &7.32&9.13&0.51\\
1.25 &120000&140.4  &7.33&8.64&0.54\\
\midrule
1.50^{\dagger} &125000&8758.5 &5.61&68.22&0.19\\
1.50 &125000&136.9  &7.41&8.53&0.46\\
1.50 &122500&123.4  &7.42&8.10&0.54\\
1.50 &120000&111.1  &7.43&7.68&0.66\\
\midrule
\end{tabularx}
\end{table}

There is an alternative method to determine the distance to H4-1 based on 
the adopted spectro-photometry data. 
We estimate the distance by comparing the luminosity derived from the sum of the atomic 
gas/molecular H$_{2}$ line emission and dust continuum emission with the theoretical post-AGB 
evolution models by \citet{2016A&A...588A..25M}. 
Detailed explanations as to how to derive the distance are found in Appendix\,\ref{S-Appen}. 
Here, we consider 0.85, 1.25, and 1.5\,M$_{\sun}$ initial-mass stars following the discussion given in the previous section.
Based on the evolutionary tracks by \citet{2016A&A...588A..25M},
0.85, 1.25, and 1.5\,M$_{\sun}$ initial-mass stars are expected to become 
white dwarfs with the final core-mass of 0.53, 0.58, and 0.60\,M$_{\sun}$, respectively.
Hence, the total ejected mass during the evolution of the 0.85, 1.25, and 1.5\,M$_{\sun}$ progenitor star is
0.32, 0.67, and 0.90\,M$_{\sun}$, respectively.

In optically thick conditions,  
the distance ($D$) to a PN can be recovered if the luminosity of the central star ($L_{\ast}$) is known,
as the total luminosity should equal to the integrated flux -- the sum of nebular continuum and lines from the ionized and atomic gas, molecular lines (i.e., H$_{2}$ and PAHs), and dust continuum -- multiplied by $4{\pi}D^{2}$. Under such optically thick conditions,
we can simultaneously fit $D$, $L_{\ast}$, and the surface gravity ($\log_{10}\,g$) of the central star 
for different initial masses at different $T_{\rm eff}$ (representing different evolutionary epochs)
as summarized in table\,\ref{T-dis}. 
With this method, the uncertainty of $D$ comes out to be $\sim10$\,$\%$ at 95\,$\%$ confidence.

For each star of specific initial mass in table\,\ref{T-dis}, 
the first row corresponds to an evolutionary epoch 
while the central star is still becoming hot prior to the PN turn-off,
whereas the following three rows correspond to the subsequent epochs along the cooling track after the PN turn-off.
While the central star is still getting hotter as it evolves toward the PN turn-off, 
the ejected gas mass can be estimated using the relation $m_{\rm g}$ = $(1.28 \pm 0.24)\times10^{-3}$\,$D_{\rm kpc}^{2}$\,M$_{\sun}$ established in section\,\ref{S-gdr}: 
a star of 0.85\,M$_{\sun}$, 1.25\,M$_{\sun}$, and 1.5\,M$_{\sun}$ initial-mass 
(at $D=25.03$, $62.21$, and $68.22$\,kpc)
would have to have ejected the circumstellar mass of $\ge0.83$\,M$_{\sun}$, $\ge5.15$\,M$_{\sun}$, and $\ge6.19$\,M$_{\sun}$,
respectively. 
Since these ejecta masses are obviously too large, 
H4-1 must be a PN presently on the white dwarf cooling track. 
If so, H4-1 is unlikely to be a $\sim0.8$\,M$_{\sun}$ initial-mass star 
unless its evolution were accelerated.
Therefore, the distance toward H4-1,
presently being a $1.25-1.5$\,M$_{\odot}$ white dwarf on the cooling track, 
would be $\sim10$\,kpc or less (table\,\ref{T-dis}).

\subsection{The circumstellar gas mass \label{S-bin}}

Now we want to consider the circumstellar mass.
If the detected gas mass is more than $0.32\,\mathrm{M}_{\sun}$, 
the progenitor would be most likely a binary system.
Thus, we need to verify the binary nature of H4-1 by estimating the amount of the circumstellar mass-loss ejecta.
To do so, we construct photoionization models using {\sc Cloudy} (v17.03; \citealt{2017RMxAA..53..385F})
by adopting the distances derived above.

\begin{table*}
\caption{
\label{T-model}
The adopted and derived parameters in the {\sc Cloudy} models.
GDR1 value corresponds to the ratio of the total gas mass to the AC mass.
GDR2 value corresponds to the ratio of the total gas mass to the (AC+PAH) mass.
}
\begin{tabularx}{\textwidth}{@{\extracolsep{\fill}}l@{\hspace{12pt}}ll}
\midrule
Central star&Values for 1.25\,M$_{\sun}$ model &Values for 1.50\,M$_{\sun}$ model\\ 
\midrule
$T_{\rm eff}$ / $\log_{10}\,g$ / $L_{\ast}$ / $D$ &127452\,K/7.30\,cm\,s$^{-2}$/ 192\,$L_{\sun}$/ 10.117\,kpc
                                        &128428\,K/7.40\,cm\,s$^{-2}$/160\,$L_{\sun}$/9.225\,kpc\\
\midrule
Nebula & Values for 1.25\,M$_{\sun}$ model& Values for 1.50\,M$_{\sun}$ model\\
\midrule 
Elemental abundances            &He: 11.04, C: 8.79,  N: 7.52, O: 8.20, &He: 11.03, C: 8.79, N: 7.50, O: 8.20,\\
(12 + $\log_{10}$$n$(X)/$n$(H)) &Ne:  6.28, S: 5.07, Ar: 4.44           &Ne:  6.25, S: 5.08, Ar: 4.46 \\ 
                 &Others: \citet{2012ApJ...747....2L}      &Others: \citet{2012ApJ...747....2L}\\
Geometry         &spherical shell nebula                   &spherical shell nebula\\
                 &inner radius = 2399\,AU (0.24{\arcsec})  &inner radius = 2359\,AU (0.20{\arcsec})\\
                 &outer radius = 23356\,AU (2.31{\arcsec}) &outer radius = 30313\,AU (2.57{\arcsec})\\
Ionization boundary radius&13252\,AU (1.31{\arcsec})             &12038\,AU (1.31{\arcsec})\\
Radial hydrogen density profile  &1297 cm$^{-3}$ in $r \le 13252$\,AU      &1062 cm$^{-3}$ in $r \le 12038$\,AU\\
 ($n_{\rm H}(r)$)                &2722 cm$^{-3}$ in $r > 13252$\,AU        &2037 cm$^{-3}$ in $r > 12038$\,AU\\
Deredden {\hb} line flux ($I$({\hb}))       & $4.54\times10^{-13}$\,erg\,s$^{-1}$\,cm$^{-2}$ & $4.54\times10^{-13}$\,erg\,s$^{-1}$\,cm$^{-2}$\\
Temperature floor      &727\,K                                   &749\,K\\
H$_{2}$ column density ($\log_{10}N({\rm H_{2}})$ &18.69\,cm$^{-2}$                 &18.61\,cm$^{-2}$\\
Ionized gas mass &0.194\,M$_{\sun}$                        &0.169\,M$_{\sun}$\\
Neutral gas mass &0.186\,M$_{\sun}$                        &0.208\,M$_{\sun}$\\ 
Total gas mass   &0.380\,M$_{\sun}$                        &0.377\,M$_{\sun}$\\
\midrule
AC and PAHs         &Values for 1.25\,M$_{\sun}$ model          &Values for 1.50\,M$_{\sun}$ model\\
\midrule 
AC  temp.           &$25 - 122$\,K                         &$24 - 131$\,K\\
PAH temp.           &$83 - 198$\,K                         &$80 - 210$\,K\\ 
AC mass             &$9.31{\times}10^{-4}$\,M$_{\sun}$     &$8.78{\times}10^{-4}$\,M$_{\sun}$\\
Ionized PAH mass    &$3.37{\times}10^{-5}$\,M$_{\sun}$     &$3.12{\times}10^{-5}$\,M$_{\sun}$\\
Neutral PAH mass    &$1.82{\times}10^{-5}$\,M$_{\sun}$     &$1.95{\times}10^{-5}$\,M$_{\sun}$\\
AC + PAH masses       &$9.83{\times}10^{-4}$\,M$_{\sun}$     &$9.29{\times}10^{-4}$\,M$_{\sun}$\\
GDR1                &405                                 &429\\
GDR2                &386                                 &406\\ 
\midrule
Reduced-$\chi^{2}$ &15.1 &17.8\\
\midrule
\end{tabularx}
\end{table*}

The central star is supposedly on the white dwarf cooling track, 
and its large surface gravity suggests that heavy elements remain deep in the photosphere. 
This means that the surface stellar abundances would be composed mainly of light elements. 
Hence, as the input SED, we adopt theoretical spectra of non local thermodynamic equilibrium (non-LTE) stellar 
atmosphere models by \citet{Rauch:2003aa} for stars whose atmosphere is composed of 
He and H (the He/H mass ratio of 0.35). 

Metals are significant coolants in ionized nebula.
With {\sc Cloudy}, we adopt elements up to Zinc as coolants. 
Starting from the observed elemental abundances (table\,\ref{T-abund}) as the input values,
we converge on the best-fit model by iteratively adjusting elemental abundances.
To constrain carbon abundance, we use the CEL C lines. 
The abundances of the undetected elements lighter than Zn are fixed 
at the values predicted for models of 1.25\,M$_{\sun}$ and 1.5\,M$_{\sun}$ initial masses \citep{2012ApJ...747....2L}.

To estimate the amount of dust, we consider spherical amorphous carbon particles 
of $0.01-0.50$\,{\micron} radius ($a$) following the power-law size distribution (of $a^{-3.5}$). 
The best-fit grain size range is determined based on a grid of models.
The broad PAH band at $5-9$\,{\micron} and two emission bands 
at 11.3/12.7\,{\micron} are attributed to charged and neutral species, respectively. 
Hence, we adopt both charged and neutral PAH molecules with the radius of $(3.6-11)\times10^{-4}$\,{\micron} that follow the same $a^{-3.5}$ size distribution. 
In computing PAH emission, we adopt the optical constants by \citet{2007ApJ...657..810D}
and use the stochastic heating mechanism.

We aim at reproducing pure-rotational H$_{2}$ lines at 9.67, 12.27, and 17.03\,{\micron} 
as well as all the detected atomic lines. 
As discussed by \citet{Otsuka:2017aa} and \citet{Otsuka:2020aa}, 
the presence of H$_{2}$ lines in the mid-IR requires 
high hydrogen densities in a warm temperature photodissociation region (PDR).
Thus, we adopt a two-layer spherical nebula with distinct hydrogen density radial profiles 
($n_{\rm H}(r)$, where $r$ is the distance from the central ionization source) to be fit through a grid of models. We set the lower temperature limit in PDRs using the temperature floor option. 
We can regard the value of the temperature floor as the excitation temperature for these three 
H$_{2}$ lines. We adopt the UMIST chemical reaction network by \citet{McElroy:2013aa}.

\begin{figure*}
\includegraphics[width=\textwidth]{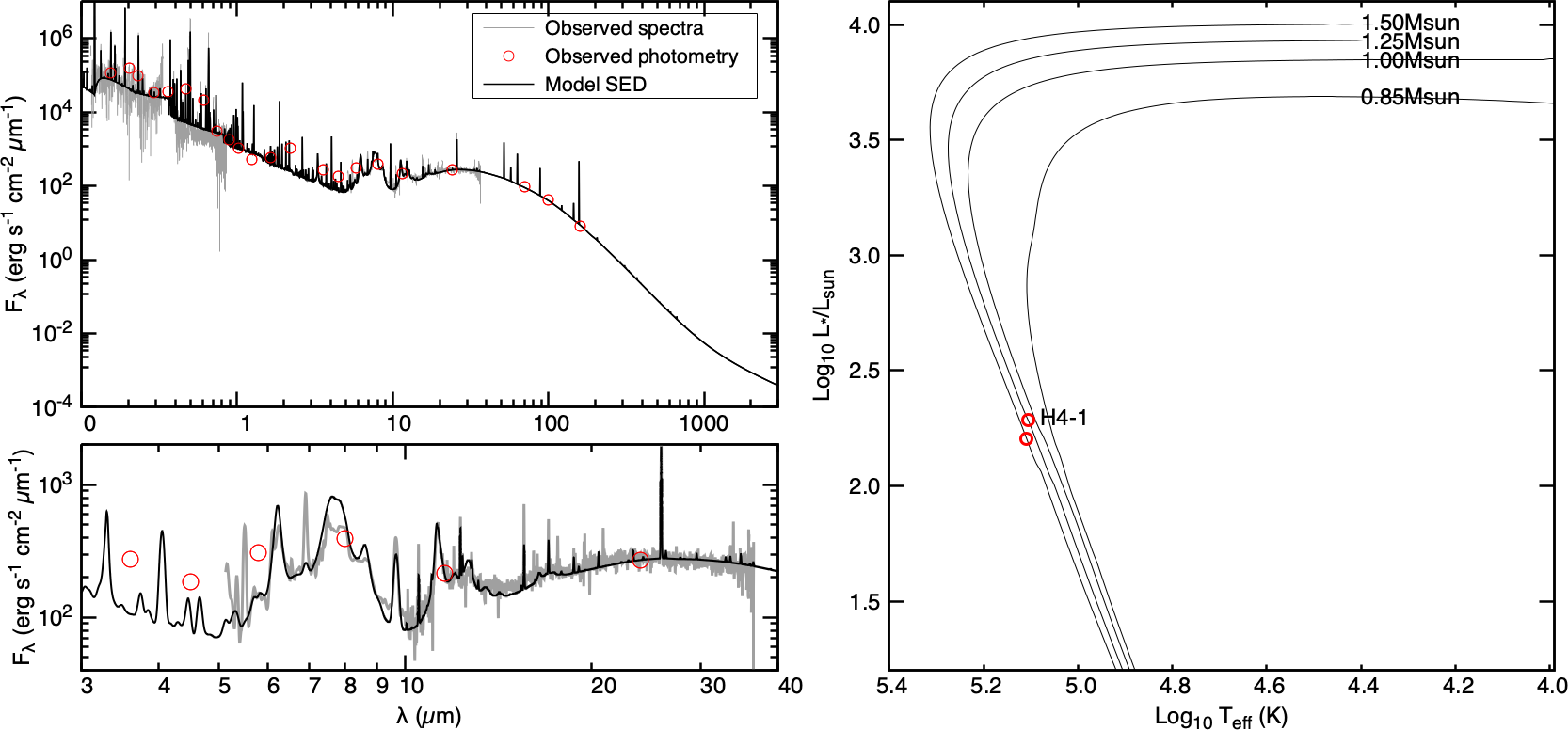}
\caption{({\it left two panels}) Comparison between the observed spectra/photometry and the predicted 
SED for the 1.25\,M$_{\sun}$ star. The lower panel focuses on the $3-40$\,{\micron} range, where 
the spectral resolution of the model SED is matched with the {\it Spitzer}/IRS spectrum. 
({\it right panel}) The current location of the central star plotted on the theoretical post-AGB 
evolutionary tracks by \citet{2016A&A...588A..25M}. 
\label{F-sed2}
}
\end{figure*}

We vary $T_{\rm eff}$ between 120000 and 130000\,K. 
At each $T_{\rm eff}$, we calculate $L_{\ast}$, $\log_{10}\,g$, 
and $D$ by the method described in subsection\,\ref{S-dist}, and then, 
keep $T_{\rm eff}$, $L_{\ast}$, $\log_{10}\,g$, and $D$. 
We emphasize that $T_{\rm eff}$, $L_{\ast}$, $\log_{10}\,g$ values for each initial mass star 
are completely along to those predicted in the post-AGB evolutionary tracks of \citet{2016A&A...588A..25M} and also 
$L_{\ast}$, $\log_{10}\,g$, and $D$ are the function of $T_{\rm eff}$. 
We vary the following 12 parameters: 
7 elemental abundances (He/C/N/O/Ne/S/Ar), the inner radius of the nebula, 
temperature floor in PDRs, GDR determined from neutral/ionized PAHs and amorphous carbon mass fractions. 
We vary the temperature floor in order to reproduce the selected H$_{2}$ lines.
Iterative calculations are terminated either when the predicted flux density at 100\,{\micron} 
reaches the observed flux in the \emph{Herschel}/PACS 100\,{\micron} band or when 
the predicted total gas mass amounts to 0.9\,M$_{\sun}$. 
The quality of fit is based on the reduced-$\chi^{2}$ calculated from the following 78 constraints:
47 atomic and 3 H$_{2}$ line fluxes relative to $I$({\hb}), 
$\log_{10}$\,$I$({\hb}),  
3 PAH band fluxes,        
23 mid- and far-IR flux densities, 
and the ionization boundary radius of 1.31{\arcsec}\footnote{This ionization boundary radius corresponds to 1.3 times Gaussian FWHM of the {\hb} emission image, within which the 85\,$\%$ of the total flux is accounted for.}, 
which is defined as the radial distance from the central star at which {\te} drops below 4000\,K.

Table\,\ref{T-model} summarizes all the input and derived/best-fit quantities for each of the two distinct initial mass models. 
The distance in the best-fit for 1.25 and 1.50\,M$_{\sun}$ stars is 10.12 and 9.23\,kpc, respectively. 
The goodness of fit according to the reduced-$\chi^{2}$ ($>15$) is dictated by the two {\oii} lines. Excluding these lines, the reduced-$\chi^{2}$ value diminishes to $\sim7$, 
which is comparable to other similarly detailed {\sc Cloudy} model results for two PNe, 
NGC650-1 \citep{2013A&A...560A...7V} and NGC6781 \citep{Otsuka:2017aa}.
Hence, we can say that both 1.25 and 1.5\,M$_{\sun}$ models equally reproduce the observational results.
Table\,\ref{T-CLOUDY2} presents a full comparison between 
the observed and modeled values, including atomic and molecular hydrogen line fluxes, PAH band fluxes at $6-9$\,{\micron} and 11\,{\micron}, fluxes for two bands near the peak dust emission intensity wavelength, and the dust emission flux density at 23 wavelengths. 
Figure\,\ref{F-sed2} shows comparisons of the SED between the observed data and 1.25\,M$_{\sun}$ initial mass model results
(left panels) as well as where each of the two models is found on the evolutionary tracks (\citealt{2016A&A...588A..25M}; right panel)
Both models successfully reproduce the observed SED across all wavelengths.
Also, the observed GDR ($387 \pm 86$) is reproduced fairly well in the models (386 and 406 in table\,\ref{T-model}). 
Hence, we conclude that the models represent H4-1 fairly well.
We emphasize here that both the ionized and neutral gas components are separately accounted for in the models, residing inside and outside of the ionization boundary, respectively. 
We immediately see that H4-1 is surrounded by as much neutral gas as ionized gas.
Hence, one must miss the whole picture of a PN if only the ionized gas component is focused on.

Both models suggest the total gas mass of $0.38 \pm 0.08$\,M$_{\sun}$ at 
95\,\% confidence, which exceeds 
the upper and lower limit of the circumstellar gas mass for the single star and binary system scenario,
respectively (subsection\,\ref{S-bin}).
Given the sensitivity of observations and 
how the circumstellar shell surface brightness drops off as a function of radial distance from the central star,
combined with our previous experience in modeling PNe, the neutral gas mass estimated here accounts for probably $\sim30-50\,\%$ of the existing neutral gas mass \citep[e.g.,][]{Otsuka:2017aa,Otsuka:2020aa}. 
This means that there is a sensitivity limit beyond which tracing the mass-loss history in the earlier AGB phase is challenging.
Based on these quantities, the total gas mass is scaled to be $0.57-0.81\,\mathrm{M}_{\sun}$ 
and $0.59-0.86\,\mathrm{M}_{\sun}$ 
for the 1.25\,M$_{\sun}$ and 1.50\,M$_{\sun}$ initial-mass models, respectively. 
Accordingly, the models suggest that the initial mass of the progenitor 
(i.e., the sum of the present mass of the central star and the total circumstellar gas mass)
is $0.58 + (0.57 \mathrm{~to~} 0.81) = 1.15\mathrm{~to~}1.39$\,M$_{\sun}$
and $0.60 + (0.59 \mathrm{~to~} 0.86) = 1.19\mathrm{~to~}1.46$\,M$_{\sun}$
for the 1.25\,M$_{\sun}$ and 1.50\,M$_{\sun}$ initial-mass models, respectively,
resulting in very self-consistent. From the argument made above, we can confidently 
conclude that the binary system scenario is more favorable for H4-1.

\subsection{The evolutionary history of H4-1 as a binary
\label{S-binmodel}}

Provided that the progenitor binary system of H4-1 now evolves as a single $\sim1.25-1.50$\,M$_{\sun}$
after a merger event between a low-mass main-sequence secondary of $\sim0.8$\,M$_{\sun}$ initial mass
and an evolved primary,
we would like to understand the initial mass of the primary of the progenitor binary system.
We also expect that binary models can explain the observed N, O, and Ne abundances unexplained by AGB 
nucleosynthesis models for single stars.

\begin{table*}
\caption{Comparison between the elemental abundances predicted by the 
binary nucleosynthesis models and by our observation. 
The second row indicates the initial abundances of 
the primary and secondary stars, which are consistent with the values predicted for $Z=10^{-4}$ 
by the Galaxy chemical evolution models by \citet{2011MNRAS.414.3231K}.
}
\centering
\begin{tabularx}{\textwidth}{@{\extracolsep{\fill}}lccccccc}
\midrule
      & He   & C    & N    &O     &Ne    & S    & Ar   \\
\midrule
Binary model &11.04 & 9.03 & 7.47 & 7.87 & 6.54 & 5.35 &4.59 \\
initial abund. &10.90 & 6.06 & 4.64 & 6.82 & 6.06 & 5.26 &4.49 \\
\midrule
H4-1         &11.02 & 8.86 & 7.74 & 8.19 & 6.42 & 5.14 & 4.61 \\
\midrule
\end{tabularx}
\label{T-binmodel}
\end{table*}

For that purpose, we run a large grid of binary models using the {\sc binary-c} code 
\citep[][ver 2.2.3]{2004MNRAS.350..407I}, for which the parameters varied are the initial mass of 
the primary ($M1$) and the secondary ($M2$) and the initial separation ($R_{\rm sep}$). 
The adopted mass range for $M1$ and $M2$ is $1.10-1.89$\,M$_{\sun}$ and $0.80-0.90$\,M$_{\sun}$, respectively, 
with a constant step of 0.01\,M$_{\sun}$. 
The range of $R_{\rm sep}$ is varied from 150\,$R_{\sun}$ to 500\,$R_{\sun}$ with a constant step of 1\,$R_{\sun}$. 
To be consistent with the Galaxy chemical evolution model by \citet{2011MNRAS.414.3231K}, 
we adopt $\mathrm{[}\alpha\mathrm{/Fe]} = +0.3$. 
The PMZ mass associated with the He-intershell products is set to be 0, which is in line with the discussion above in subsection\,\ref{S-dis-ori}. Overshooting is not considered. 
We adopt the Reimer's mass-loss law with the fitting parameter $\eta=0.5$ during the red giant stage,
and specific rates of mass loss during the early AGB phase \citep[we set GDR = 400 from our study;][]{2017MNRAS.465..403G} 
and the thermal pulsing AGB phase \citep{1993ApJ...413..641V}.
Note that the models by \citet{2012ApJ...747....2L} referenced earlier adopt specific rates of mass loss \citep{1993ApJ...413..641V} throughout the AGB phase irrespective of thermal pulses.  
Adopting high mass loss during the early AGB phase \citep{2017MNRAS.465..403G} can suppress the unwanted Ne enhancement, compared with an exclusive use of the rates of mass loss during the thermal pulsing AGB phase \citep{1993ApJ...413..641V}. This is because that 
the mass loss in the early AGB determines the mass in the beginning of TP phase and 
affects the duration of the TP phase. 
We set the orbital eccentricity of 0, common envelope efficiency ($\alpha$)
of 0.5, and structure parameter ($\lambda$) of 0.5. $\alpha$ is 
the efficiency of converting the orbital energy into the kinetic energy of the envelope 
and $\lambda$ depends on the structure of the primary star, conventionally chosen as a 
constant, typically $\sim0.5$ \citep{2014MNRAS.442.1980Z}. $\alpha$ and $\lambda$ 
control the ratio of final to initial orbital separation before and after common envelope phase. 
The remaining parameters required to simulate the binary evolution are set to 
the default values in the {\sc Binary-c} code. 
Out of the total number of models that turns out to be 308880, 
favorable candidates are selected as those satisfying the following criteria; 
(1) the total duration of evolution exceeds 10\,Gyr, 
(2) the final mass of the primary is greater than 0.58\,M$_{\sun}$, and 
(3) the C/O ratio (by number) is $> 1$. 
Among these favorable candidates, we find out the best-fit model for the progenitor system of H4-1 
by evaluating the reduced-$\chi^{2}$ between the observed and predicted He/C/N/O/Ne/S/Ar abundances,
assuming that the uncertainty of each elemental abundance is of 0.1\,dex.

In table\,\ref{T-binmodel}, we compare the best-fit model abundances (the top row) with the observed abundances (the bottom row), while the initial abundances adopted for the models are shown in the middle. Adopting the observed C abundance including the C atom incorporated into the dust grains ($9.02 \pm 0.11$\,dex), 
the binary model gives better fitting. Either way, the binary model fairly reproduces all the observed abundances. 
The best-fit model (the reduced-$\chi^{2}$ of 1.58) suggests 
the initial mass of 1.87\,M$_{\sun}$ and 0.82\,M$_{\sun}$ for the primary and secondary, respectively,
with the initial separation of 248\,$R_{\sun}$. 
Presently, the now-merged star is of 0.62\,M$_{\sun}$ on the white dwarf cooling track at the age of 10.02\,Gyr since the formation of the progenitor binary system in the Galactic halo.

The evolutionary history of H4-1, according to the best-fit model, is as follows. 
About 3.8\,Gyrs after the Big Bang, 
a Pop II binary system was formed in the young Milky Way.
When the primary of 1.87\,M$_{\sun}$ initial mass became an early AGB star, 
the primary began to lose its envelope to the Roche-lobe overflow, rapidly evolving to be a CO white dwarf of 
$\sim0.6$\,M$_{\sun}$ without experiencing TPs and TDUs. 
As the secondary went into the giant phases, it
experienced the Roche-lobe overflow several times 
while never losing its outer envelope.
The surface of the secondary was 
contaminated by the primary, its He/C/N/O/Ne/S/Ar composition was changed into 10.97/7.65/7.43/7.77/6.33/5.33/4.57\,dex, respectively. 
The N-excess observed in H4-1 can be explained by the result of mass-transfer from the primary to the secondary stars.
After about 10\,Gyrs since the formation, the back and forth large-scale mass transfer between the white dwarf primary and the main sequence secondary via the Roche-lobe overflow was 
repeated several times, while the binary separation was reduced each time. 
Ultimately, the system became a contact binary and then coalesced into a single star of 1.42\,M$_{\sun}$,
showing the surface composition identical to the sum of those from both stars.  
 The merged star subsequently underwent the TPs and TDUs required to become C-rich during the AGB phase of its own and became a CO white dwarf of 0.62\,M$_{\sun}$ surrounded by a C- and N-rich and Ne-poor dusty nebulae. The multipolar structure of the nebula and the presence of the equatorial torus are consistent with the binarity of the progenitor system.
 
If adopting mass loss by \citet{1993ApJ...413..641V} through the AGB phase, the best-fit model 
(the 1.65\,M$_{\sun}$ primary and 0.86\,M$_{\sun}$ secondary stars 
with the initial separation of 274\,$R_{\sun}$) predicts the binary system ended 
as a 0.64\,M$_{\sun}$ CO white dwarf after binary mass-transfer and merger, but high C and Ne abundances\footnote{The model predicts that He/C/N/O/Ne/S/Ar 
composition is 11.00/9.24/7.45/7.88/7.23/5.34/4.58\,dex, respectively.}. The low Ne in H4-1 
would be due to the high mass loss during the early AGB. 

As we discussed above, binary evolution completely resolved the long-standing issues on 
evolutionary time scale and anormal chemical abundance pattern. 
Therefore, we confidently conclude that H4-1 originated from a Pop II binary star.

\section{Summary
\label{S-sum}}

We have performed a comprehensive study of the Galactic halo PN H4-1 
based on the newly acquired Seimei/KOOLS-IFU data 
and auxiliary multiband spectral photometry data 
to reveal its origin and evolution through accurate measurements of gas abundances and mass.
The spatially-resolved emission line maps reconstructed from the KOOLS-IFU data cube exhibit both 
the elliptical elongation along the polar axis and the central flattening along the equatorial 
plane commonly seen as a signature of the presence of an equatorial disk structure in bipolar PNe evolved from high-mass progenitors. 

Fully data-driven analyses of the spectra that are point-spread-function-matched at each wavelength bin have yielded seven elemental abundances, gas-to-dust mass ratio, and gas/dust masses, 
which are also supported by the distance to the object newly estimated consistently by ourselves.
The detailed photoionization model has reproduced all the observed quantities of the adopted multiband spectral photometry data. 
The derived total gas mass strongly indicates 
that the progenitor is not a typical single halo star but a binary system. 
By modeling the possible binary evolution and two-star nucleosynthesis, 
we have been able to reproduce the observed elemental abundances 
that cannot be explained by nucleosynthesis of a single star.
We have also successfully 
estimated the initial mass and the initial orbital separation of the progenitor binary system. 
The best-fit model predicts that 
the progenitors of 1.87\,M$_{\sun}$ and 0.82\,M$_{\sun}$ initial mass were originally Pop II stars 
formed about 4\,Gyrs after the Big Bang, which underwent episodes of mass-transfers 
and an eventual binary merger, subsequently evolved into a PN showing unique chemical abundances. 
The N-excess would be due to a result of mass-transfer, and the extremely low Ne would be due to the high mass loss during the early AGB. 
Mutual follow-up between halo PNe and CEMP stars will greatly help in understanding
their origin and evolution.

\begin{ack}
We are grateful to the referee for a careful reading and valuable suggestions. 
MO thanks Dr. Maria Lugaro for fruitful discussion on chemical abundances in metal-poor stars. 
MO thanks Drs. Robert Izzard and David Hendriks to kindly instruct him about the use of their developed codes. MO was supported by the Japan Society for the Promotion of Science (JSPS) Grants-in-Aids for Scientific Research (C) (19K03914 and 22K03675). 
MO thanks his parents for their deep love and support. 
TU was partially supported by the JSPS through its invitation fellowship program (FY2020, long-term).
\end{ack}

\bibliographystyle{apj}

\appendix 
\section{The method to calculate the nebula radiation}

\setcounter{table}{0}
\renewcommand{\thetable}{A\arabic{table}}

\setcounter{figure}{0}
\renewcommand{\thefigure}{A\arabic{figure}}

\label{S-Appen}

\begin{figure}
\includegraphics[width=\columnwidth]{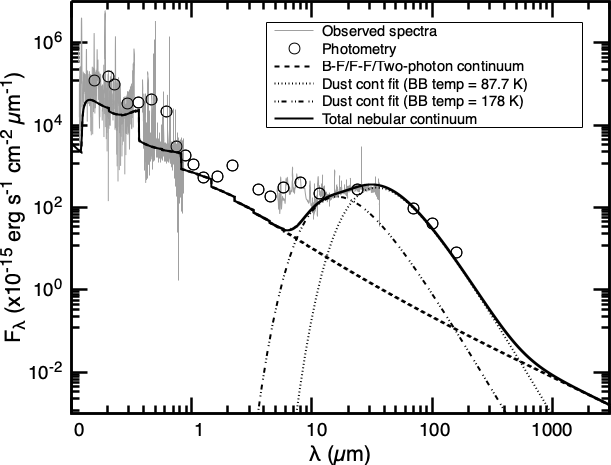}
\caption{
The spectral energy distribution (SED) plots are generated from the observed dereddened spectra and photometry, shown as gray lines and circles respectively. The broken line represents the sum of the {\hi}/{\hei}/{\heii} bound-free (B-F), free-free (F-F), and two-photon continuum calculated using {\sc Nebcont} in {\sc Starlink} \citep{2014ASPC..485..391C}. We fitted the dust continuum in the \emph{gas-emission-free} \emph{Spitzer}/IRS spectrum and \emph{Herschel}/PACS 70/100/160\,{\micron} photometry SED with a modified  two-temperature (148 and 69.5\,K) blackbody function (dashed and dot lines, respectively). 
The total nebular continuum, indicated by the black line, is the sum of the dust continuum fit and the B-F/F-F/two-photon continuum.}
\label{F-sed}
\end{figure}

We calculated the nebular radiation by following steps. 
First, we calculated the total line flux from the observed atomic and molecular H$_{2}$ lines. The expected fluxes of {\hi} and {\heii} lines appeared outside wavelengths in the obtained spectra are calculated using {\hi} and {\heii} recombination coefficients in {\te} = 12634\,K and {\Ne} = 420\,cm$^{-3}$ for {\hi} and {\te}({\oiii}) and {\Ne}({\siii}) for {\heii}. The expected CEL fluxes are calculated by solving atomic models of each ion in the 
obtained {\te}({\oiii}, {\nii}) and {\Ne}({\sii}, {\siii}). 
The following atomic lines are considered: {\hi} 
(Lyman\,$\alpha$; 
Balmer $n-2$, $n=3-40$, $n$: quantum number; 
Paschen $n-3$, $n=4-40$; 
Bracket $n-4$, $n=5-40$; 
Pfund $n-5$, $n=6-40$), 
He\,{\sc ii} 
(Fowler $n-3$, $n=4-50$; 
Pickering $n-4$, $n=5-40$),  CEL $[${\cii}$]$\,157\,{\micron}, {\nii}\,121/205\,{\micron}, {\oi}\,63/145\,{\micron}, {\oiii}\,51/88\,{\micron}, {\oii}\,3726/29\,{\AA}, {\neiii}\,3868/3967\,{\AA}. Thus, we obtained the total atomic and H$_{2}$ line flux of $3.41\times10^{-11}$\,{\ergsF} by simply summing up the observed and expected fluxes of the atomic and molecular H$_{2}$ lines. 

Second, we generated the integrated spectral energy distribution (SED) of the {\hi}/{\hei}/{\heii} bound-free, free-free, and two-photon continuum using {\sc Nebcont} in {\sc Starlink} \citep{2014ASPC..485..391C} by adopting 
$I$({\hb}) = $4.54\times10^{-13}$\,{\ergsF}, 
{\te} = 12634\,K, {\Ne} = 420\,cm$^{-3}$, and 
He$^{+,2+}$ abundances (table\,\ref{T-abund}). 

Third, we fit a two temperature (149 and 69.5\,K) modified blackbody function to the dust continuum in the range from 14 to 160\,{\micron} based on the {\it gas-emission-free} \emph{Spitzer}/IRS spectrum and the \emph{Herschel}/PACS photometry SED. Here, we adopt $Q_{\rm ext}$ in the case of 
amorphous carbon grain (i.e., $Q_{\rm ext } \propto {\lambda}^{-1}$). 
The expected flux contributions of the gas emission to each 70/100/160\,{\micron} band were calculated 
by taking the PACS throughput at each wavelength 
into account. The {\it gas-emission-free} PACS 70/100/160\,{\micron} band flux densities are 
$9.65\times10^{-14}$, 
$3.99\times10^{-14}$, and 
$7.44\times10^{-15}$\,{\ergsm}, respectively. 
Then, we added the resultant fit to the nebular {\hi}/{\hei}/{\heii} continuum calculated in the second step. 

Thus, we obtained a synthesized nebular continuum SED (figure\,\ref{F-sed}). 
For the continuum flux in $5.2-36.0$\,{\micron}, we used the \emph{Spitzer}/IRS spectrum where the 
gas emission was removed out but PAHs were not. Finally, we obtained the nebular continuum-flux integrated between 0.1 and 600\,{\micron} to be $2.61\times10^{-11}$\,{\ergsF}.
The breakdown of the integrated nebular-line and continuum fluxes is summarized in table\,\ref{T-conti}.

\section{Supporting Tables}

The follwings are the supproting tables.

\begin{table*}
\caption{Broadband flux densities of H4-1. 
\label{T-obsphot}}
\scriptsize
\begin{tabularx}{\textwidth}{@{\extracolsep{\fill}}D{.}{.}{-1}D{.}{.}{-1}D{p}{\pm}{-1}D{p}{\pm}{-1}cc}
\midrule
\multicolumn{1}{r}{$\lambda_{\rm c}$~~~} & 
\multicolumn{1}{r}{$f$($\lambda$)~~~} & 
\multicolumn{1}{c}{$F_{\lambda}$} & 
\multicolumn{1}{c}{$I_{\lambda}$} & 
Sources & 
Ref.\\
\multicolumn{1}{r}{({\micron})} & 
&
\multicolumn{1}{c}{(erg\,s$^{-1}$\,cm$^{-2}$\,{\micron}$^{-1}$)}&
\multicolumn{1}{c}{(erg\,s$^{-1}$\,cm$^{-2}$\,{\micron}$^{-1}$)}\\
\midrule
0.1549 & ~1.2383 & 6.95\times10^{-11} ~p~ 4.04\times10^{-13} & 2.02\times10^{-10} ~p~ 6.94\times10^{-11} & GALEX & (1)\\ 
0.2042 & ~1.5521 & 8.24\times10^{-11} ~p~ 1.09\times10^{-12} & 2.79\times10^{-10} ~p~ 1.09\times10^{-10} & XMM & (2) \\ 
0.2305 & ~1.4439 & 5.35\times10^{-11} ~p~ 1.49\times10^{-13} & 1.72\times10^{-10} ~p~ 6.44\times10^{-11} & GALEX & (1)\\ 
0.2934 & ~0.5938 & 2.28\times10^{-11} ~p~ 2.12\times10^{-13} & 4.88\times10^{-11} ~p~ 1.19\times10^{-11} & XMM &  (2)\\ 
0.3608 & ~0.3455 & 2.61\times10^{-11} ~p~ 2.21\times10^{-14} & 4.96\times10^{-11} ~p~ 1.02\times10^{-11} & SDSS & (3) \\ 
0.4672 & ~0.0538 & 3.28\times10^{-11} ~p~ 1.47\times10^{-14} & 5.42\times10^{-11} ~p~ 8.76\times10^{-12} & SDSS & (3) \\ 
0.6141 & -0.2409 & 1.76\times10^{-11} ~p~ 7.84\times10^{-15} & 2.53\times10^{-11} ~p~ 2.95\times10^{-12} & SDSS & (3) \\ 
0.7458 & -0.4168 & 2.68\times10^{-12} ~p~ 2.16\times10^{-15} & 3.54\times10^{-12} ~p~ 3.16\times10^{-13} & SDSS & (3) \\ 
0.8923 & -0.5828 & 1.64\times10^{-12} ~p~ 3.72\times10^{-15} & 2.00\times10^{-12} ~p~ 1.28\times10^{-13} & SDSS & (3) \\ 
1.0305 & -0.6694 & 9.99\times10^{-13} ~p~ 1.63\times10^{-15} & 1.17\times10^{-12} ~p~ 5.93\times10^{-14} & UKIDSS & (4) \\ 
1.2483 & -0.7572 & 5.05\times10^{-13} ~p~ 1.10\times10^{-15} & 5.67\times10^{-13} ~p~ 2.11\times10^{-14} & UKIDSS & (4) \\ 
1.6313 & -0.8422 & 5.56\times10^{-13} ~p~ 2.50\times10^{-15} & 6.00\times10^{-13} ~p~ 1.48\times10^{-14} & UKIDSS & (4) \\ 
2.2010 & -0.9026 & 1.01\times10^{-12} ~p~ 2.93\times10^{-15} & 1.06\times10^{-12} ~p~ 1.61\times10^{-14} & UKIDSS & (4) \\ 
3.6000 & -0.9750 & 2.72\times10^{-13} ~p~ 2.14\times10^{-14} & 2.75\times10^{-13} ~p~ 2.16\times10^{-14} & \emph{Spitzer}/IRAC & (5) \\ 
4.5000 & -0.9830 & 1.85\times10^{-13} ~p~ 2.57\times10^{-14} & 1.86\times10^{-13} ~p~ 2.59\times10^{-14} & \emph{Spitzer}/IRAC & (5) \\ 
5.8000 & -0.9880 & 3.07\times10^{-13} ~p~ 6.99\times10^{-14} & 3.09\times10^{-13} ~p~ 7.03\times10^{-14} & \emph{Spitzer}/IRAC & (5) \\ 
8.0000 & -0.9840 & 3.95\times10^{-13} ~p~ 3.43\times10^{-14} & 3.98\times10^{-13} ~p~ 3.46\times10^{-14} & \emph{Spitzer}/IRAC & (5) \\ 
11.5600 & -0.9720 & 2.14\times10^{-13} ~p~ 1.68\times10^{-15} & 2.16\times10^{-13} ~p~ 1.94\times10^{-15} & WISE & (6) \\ 
24.0000 & -0.9880 & 2.68\times10^{-13} ~p~ 9.83\times10^{-15} & 2.69\times10^{-13} ~p~ 9.90\times10^{-15} & \emph{Spitzer}/MIPS & (5) \\ 
70.0000 &         & 9.66\times10^{-14} ~p~ 1.94\times10^{-15} & 9.66\times10^{-14} ~p~ 1.94\times10^{-15} & \emph{Herschel}/PACS & (7) \\ 
100.0000 &        & 4.07\times10^{-14} ~p~ 4.52\times10^{-15} & 4.07\times10^{-14} ~p~ 4.52\times10^{-15} & \emph{Herschel}/PACS & (7) \\ 
160.0000 &        & 8.14\times10^{-15} ~p~ 1.83\times10^{-15} & 8.14\times10^{-15} ~p~ 1.83\times10^{-15} & \emph{Herschel}/PACS & (7)\\
\midrule
\end{tabularx}
\begin{tabnote}
Note -- The reddening correction for the flux densities at 70, 100, and 160\,{\micron} is not performed because the 
dust extinction at these bands is negligible. The gas-emission-free PACS 70/100/160\,{\micron} band flux densities are 
$9.65\times10^{-14}$, $3.99\times10^{-14}$, and $7.44\times10^{-15}$\,{\ergsm}, respectively (see subsection\,\ref{S-dist}.)\\
References -- (1) \citet{2017ApJS..230...24B}; (2) \citet{2012MNRAS.426..903P}; (3) \citet{2015ApJS..219...12A}; 
(4) \citet{2007MNRAS.379.1599L}; (5) \citet{tajitsu:2023aa}; (6) \citet{2013wise.rept....1C}; (7) \citet{2020yCat.8106....0H}.
\end{tabnote}
\end{table*}

\begin{table*}
\caption{The detected line-fluxes normalized to the line-flux of {\hb} whose intensity is 100. \label{T-line}}
\centering
\scriptsize
\begin{tabularx}{\textwidth}{@{}@{\extracolsep{\fill}}
rD{.}{.}{-1}cD{.}{.}{-1}D{.}{.}{-1}rD{.}{.}{-1}cD{.}{.}{-1}D{.}{.}{-1}@{}}
\midrule
\multicolumn{1}{c}{$\lambda_{\rm lab}$} & \multicolumn{1}{c}{~~~$f(\lambda)$} & \multicolumn{1}{c}{Species} & \multicolumn{1}{c}{$I$($\lambda$)} & \multicolumn{1}{c}{$\delta$\,$I$($\lambda$)}
&\multicolumn{1}{c}{$\lambda_{\rm lab}$} & \multicolumn{1}{c}{~~~$f(\lambda)$} & \multicolumn{1}{c}{Species} & \multicolumn{1}{c}{$I$($\lambda$)} & \multicolumn{1}{c}{$\delta$\,$I$($\lambda$)}\\
\midrule
1335.70\,{\AA} & ~1.584 & C\,{\sc ii} & 107.376 & 27.150 & 7319/20\,{\AA} & -0.398 & {\oii} & 4.119 & 0.301 \\ 
1549.03\,{\AA} & ~1.238 & C\,{\sc iv} & 243.453 & 46.506 & 7330/31\,{\AA} & -0.400 & {\oii} & 3.514 & 0.269 \\ 
1640.03\,{\AA} & ~1.177 & {\heii} & 68.000 & 13.763 & 8467.25\,{\AA} & -0.541 & {\hi} & 0.388 & 0.101 \\ 
1664.03\,{\AA} & ~1.167 & O\,{\sc iii}$]$ & 29.920 & 8.179 & 8502.48\,{\AA} & -0.544 & {\hi} & 0.561 & 0.124 \\ 
1909.49\,{\AA} & ~1.257 & [C\,{\sc iii}]+C\,{\sc iii}$]$ & 1260.795 & 243.188 & 8545.38\,{\AA} & -0.549 & {\hi} & 0.625 & 0.122 \\ 
2321/30\,{\AA} & ~1.382 & O\,{\sc iii}$]$ & 1.817 & 0.227 & 8598.39\,{\AA} & -0.554 & {\hi} & 0.661 & 0.181 \\ 
2322-28\,{\AA} & ~1.382 & [{\cii}] & 305.116 & 66.494 & 8665.02\,{\AA} & -0.560 & {\hi} & 0.714 & 0.148 \\ 
4340.47\,{\AA} & ~0.157 & {\hi} & 44.191 & 1.545 & 8728.90\,{\AA} & -0.566 & {\cii} ? & 1.056 & 0.123 \\ 
4363.21\,{\AA} & ~0.149 & {\oiii} & 7.665 & 0.959 & 8750.47\,{\AA} & -0.568 & {\hi} & 1.182 & 0.136 \\ 
4471.50\,{\AA} & ~0.115 & {\hei} & 6.836 & 0.760 & 8862.78\,{\AA} & -0.578 & {\hi} & 1.519 & 0.178 \\ 
4685.71\,{\AA} & ~0.050 & {\heii} & 10.068 & 0.417 & 9014.91\,{\AA} & -0.590 & {\hi} & 1.620 & 0.179 \\ 
4712.42\,{\AA} & ~0.042 & {\hei}+{\ariv} & 0.770 & 0.146 & 9068.60\,{\AA} & -0.594 & {\siii} & 0.856 & 0.104 \\ 
4861.33\,{\AA} & ~0.000 & {\hi} & 100.000 & 1.121 & 9229.01\,{\AA} & -0.605 & {\hi} & 2.666 & 0.292 \\ 
4921.93\,{\AA} & -0.016 & {\hei} & 1.306 & 0.243 & 9537.45\,{\AA} & -0.625 & {\siii} & 2.296 & 0.474 \\ 
4958.91\,{\AA} & -0.026 & {\oiii} & 224.438 & 3.212 & 9545.97 \,{\AA}& -0.626 & {\hi} & 3.844 & 0.484 \\ 
5006.84\,{\AA} & -0.038 & {\oiii} & 670.451 & 9.689 & 9824.13\,{\AA} & -0.643 & [C\,{\sc i}] & 4.142 & 0.710 \\ 
5198/200\,{\AA} & -0.082 & {\NI} & 1.587 & 0.127 & 9850.26\,{\AA} & -0.644 & [C\,{\sc i}] & 14.058 & 1.571 \\ 
5411.52\,{\AA} & -0.126 & {\heii} & 0.902 & 0.116 & 10049.37\,{\AA} & -0.656 & {\hi} & 4.361 & 0.724 \\ 
5754.60\,{\AA} & -0.185 & {\nii} & 2.406 & 0.156 & 5.51\,{\micron} & -0.987 & H$_{2}$ & 5.976 & 0.917 \\ 
5875.66\,{\AA} & -0.203 & {\hei} & 12.539 & 0.488 & 6.11\,{\micron} & -0.989 & H$_{2}$ & 2.324 & 0.359 \\ 
6300.34\,{\AA} & -0.263 & {\oi}+{\siii} & 9.208 & 0.408 & 6.91\,{\micron} & -0.990 & H$_{2}$ & 10.573 & 1.621 \\ 
6363.78\,{\AA} & -0.271 & {\oi} & 3.110 & 0.153 & 9.67\,{\micron} & -0.949 & H$_{2}$ & 8.904 & 1.312 \\ 
6548.10\,{\AA} & -0.296 & {\nii} & 33.293 & 3.103 & 10.52\,{\micron} & -0.960 & {\siv} & 1.010 & 0.170 \\ 
6562.77\,{\AA} & -0.298 & {\hi} & 277.865 & 13.369 & 12.28\,{\micron} & -0.980 & H$_{2}$ & 2.254 & 0.345 \\ 
6583.50\,{\AA} & -0.300 & {\nii} & 105.668 & 5.310 & 12.40\,{\micron} & -0.980 & {\hi} & 0.871 & 0.153 \\ 
6678.16\,{\AA} & -0.313 & {\hei} & 4.274 & 0.226 & 12.81\,{\micron} & -0.983 & {\neii} & 0.325 & 0.062 \\ 
6716.44\,{\AA} & -0.318 & {\sii} & 0.957 & 0.070 & 15.56\,{\micron} & -0.985 & {\neiii} & 2.971 & 0.456 \\ 
6730.82\,{\AA} & -0.320 & {\sii} & 0.929 & 0.072 & 17.03\,{\micron} & -0.982 & H$_{2}$ & 2.070 & 0.317 \\ 
7065.25\,{\AA} & -0.364 & {\hei} & 4.873 & 0.282 & 18.71\,{\micron} & -0.981 & {\siii} & 0.764 & 0.124 \\ 
7135.80\,{\AA} & -0.374 & {\ariii} & 0.487 & 0.056 & 25.89\,{\micron} & -0.989 & {\oiv} & 25.845 & 3.930 \\ 
7236.42\,{\AA} & -0.387 & {\cii}+{\ariv} & 0.521 & 0.108 & 28.22\,{\micron} & -0.990 & H$_{2}$ & 2.043 & 0.315 \\ 
7281.35\,{\AA} & -0.393 & {\hei} & 0.769 & 0.089 &  &  &  &  &  \\ 
\midrule
\end{tabularx}
\end{table*}

\begin{table*}
\caption{Comparison between the observed and the {\sc Cloudy} model predicted values.
\label{T-CLOUDY2}
}
\scriptsize
\begin{tabularx}{\textwidth}{@{\extracolsep{\fill}}lD{.}{.}{-1}D{.}{.}{-1}D{.}{.}{-1}D{.}{.}{-1}lD{.}{.}{-1}D{.}{.}{-1}D{.}{.}{-1}D{.}{.}{-1}}
\midrule
Species & \multicolumn{1}{c}{$\lambda_{\rm lab}$ ({\AA})} & 
\multicolumn{1}{c}{$I$(Obs)} & 
\multicolumn{1}{c}{$I$(1.25\,M$_{\sun}$)} & 
\multicolumn{1}{c}{$I$(1.50\,M$_{\sun}$)} & 
Species & \multicolumn{1}{c}{$\lambda_{\rm lab}$ ({\AA})} & 
\multicolumn{1}{c}{$I$(Obs)} & 
\multicolumn{1}{c}{$I$(1.25\,M$_{\sun}$)} & 
\multicolumn{1}{c}{$I$(1.50\,M$_{\sun}$)} \\ 
\midrule
{\civ} & 1549.03 & 242.730 & 320.547 & 320.043 & {\nii} & 6583.50 & 110.303 & 107.786 & 106.932 \\ 
{\heii} & 1640.03 & 73.154 & 87.047 & 88.555 & {\hei} & 6678.16 & 4.271 & 3.619 & 3.610 \\ 
$[$C\,{\sc iii}$]$+C\,{\sc iii}$]$ & 1909.00 & 1254.077 & 1250.051 & 1228.957 & {\sii} & 6716.44 & 0.979 & 0.971 & 1.024 \\ 
{\oiii} & 2320.95 & 2.098 & 2.493 & 2.463 & {\sii} & 6730.82 & 0.948 & 1.147 & 1.214 \\ 
$[$C\,{\sc ii}$]$ & 2326.00 & 293.310 & 273.891 & 277.782 & {\ariii} & 7135.80 & 0.461 & 0.428 & 0.447 \\ 
{\heii} & 4338.55 & 0.245 & 0.310 & 0.316 & {\hei} & 7281.35 & 0.729 & 0.909 & 0.909 \\ 
{\hi} & 4340.47 & 47.394 & 47.224 & 47.218 & {\oii} & 7320.00 & 4.102 & 8.109 & 8.417 \\ 
{\oiii} & 4363.21 & 8.851 & 9.932 & 9.811 & {\oii} & 7330.00 & 3.458 & 6.490 & 6.736 \\ 
{\heii} & 4685.71 & 10.419 & 10.954 & 11.143 & {\hi} & 8467.25 & 0.393 & 0.443 & 0.442 \\ 
{\heii} & 4859.18 & 0.505 & 0.618 & 0.628 & {\hi} & 8502.48 & 0.562 & 0.521 & 0.520 \\ 
{\hi} & 4861.33 & 100.000 & 100.000 & 100.000 & {\hi} & 8545.38 & 0.664 & 0.621 & 0.620 \\ 
{\hei} & 4921.93 & 1.457 & 1.322 & 1.318 & {\hi} & 8598.39 & 0.634 & 0.751 & 0.750 \\ 
{\oiii} & 4958.91 & 217.864 & 211.798 & 208.263 & {\hi} & 8665.47 & 0.712 & 0.923 & 0.922 \\ 
{\oiii} & 5006.84 & 651.687 & 631.914 & 621.370 & {\hi} & 8750.43 & 1.233 & 1.154 & 1.152 \\ 
{\heii} & 5411.52 & 0.804 & 0.939 & 0.955 & {\hi} & 8862.78 & 1.396 & 1.472 & 1.470 \\ 
{\nii} & 5754.60 & 2.617 & 2.644 & 2.622 & {\hi} & 9014.91 & 1.578 & 1.694 & 1.692 \\ 
{\hei} & 5875.66 & 14.077 & 13.735 & 13.716 & {\siii} & 9068.60 & 0.879 & 0.774 & 0.794 \\ 
{\oi} & 6363.78 & 3.577 & 5.878 & 6.268 & {\hi} & 9229.01 & 2.568 & 2.332 & 2.328 \\ 
{\nii} & 6548.10 & 35.768 & 36.565 & 36.275 & {\siii} & 9537.45 & 2.229 & 1.911 & 1.961 \\ 
{\heii} & 6559.91 & 1.382 & 1.529 & 1.555 & $[$C\,{\sc i}$]$ & 9824.13 & 3.548 & 2.653 & 2.799 \\ 
{\hi} & 6562.77 & 278.955 & 286.983 & 287.272 & $[$C\,{\sc i}$]$ & 9850.26 & 13.766 & 7.926 & 8.363 \\ 
\midrule
Species & \multicolumn{1}{c}{$\lambda_{\rm lab}$ ({\micron})} & 
\multicolumn{1}{c}{$I$(Obs)} & 
\multicolumn{1}{c}{$I$(1.25\,M$_{\sun}$)} & 
\multicolumn{1}{c}{$I$(1.50\,M$_{\sun}$)} & 
Species & \multicolumn{1}{c}{$\lambda_{\rm lab}$ ({\micron})} & 
\multicolumn{1}{c}{$I$(Obs)} & 
\multicolumn{1}{c}{$I$(1.25\,M$_{\sun}$)} & 
\multicolumn{1}{c}{$I$(1.50\,M$_{\sun}$)} \\ 
\midrule
H$_{2}$ & 9.67 & 9.722 & 9.521 & 9.573 & {\neiii} & 15.56 & 3.272 & 2.402 & 2.225 \\ 
{\siv} & 10.52 & 1.105 & 0.921 & 0.908 & H$_{2}$ & 17.03 & 2.279 & 2.111 & 2.121 \\ 
H$_{2}$ & 12.28 & 2.479 & 2.543 & 2.573 & {\siii} & 18.71 & 0.933 & 0.936 & 0.962 \\ 
{\neii} & 12.81 & 0.358 & 0.466 & 0.499 & {\oiv} & 25.89 & 28.494 & 32.949 & 32.718 \\
\midrule
~~$F_{\nu}$ & 
\multicolumn{1}{c}{$\lambda_{\rm lab}$ ({\micron})} & 
\multicolumn{1}{c}{$F_{\nu}$(Obs)} & 
\multicolumn{1}{c}{$F_{\nu}$(1.25\,M$_{\sun}$)}  & 
\multicolumn{1}{c}{$F_{\nu}$(1.50\,M$_{\sun}$)}   & 
\multicolumn{1}{c}{$F_{\nu}$} & 
\multicolumn{1}{c}{$\lambda_{\rm lab}$ ({\micron})} & 
\multicolumn{1}{c}{$F_{\nu}$(Obs)}  & 
\multicolumn{1}{c}{$F_{\nu}$(1.25\,M$_{\sun}$)}  & 
\multicolumn{1}{c}{$F_{\nu}$(1.50\,M$_{\sun}$)}  \\ 
\midrule
IRS01 & 10.00 & 0.003 & 0.003 & 0.003 &~~IRS13  & 25.00 & 0.063 & 0.059 & 0.058 \\ 
IRS02 & 14.00 & 0.012 & 0.009 & 0.010 &~~IRS14  & 26.00 & 0.068 & 0.065 & 0.063 \\ 
IRS03 & 15.00 & 0.015 & 0.011 & 0.011 &~~IRS15  & 27.00 & 0.074 & 0.070 & 0.067 \\ 
IRS04 & 16.00 & 0.019 & 0.015 & 0.015 &~~IRS16  & 28.00 & 0.078 & 0.075 & 0.072 \\ 
IRS05 & 17.00 & 0.023 & 0.021 & 0.021 &~~IRS17  & 29.00 & 0.083 & 0.080 & 0.077 \\ 
IRS06 & 18.00 & 0.027 & 0.023 & 0.023 &~~IRS18  & 30.00 & 0.087 & 0.085 & 0.082 \\ 
IRS07 & 19.00 & 0.032 & 0.027 & 0.027 &~~IRS19  & 31.00 & 0.091 & 0.090 & 0.086 \\ 
IRS08 & 20.00 & 0.037 & 0.031 & 0.031 &~~IRS20  & 32.00 & 0.095 & 0.095 & 0.091 \\ 
IRS09 & 21.00 & 0.042 & 0.036 & 0.036 &~~IRS21  & 33.00 & 0.097 & 0.099 & 0.095 \\ 
IRS10 & 22.00 & 0.047 & 0.042 & 0.041 &~~PAC01  & 70.00 & 0.158 & 0.153 & 0.149 \\ 
IRS11 & 23.00 & 0.053 & 0.048 & 0.047 &~~PAC02  & 100.00 & 0.133 & 0.133 & 0.134 \\ 
IRS12 & 24.00 & 0.058 & 0.054 & 0.052 &~~PAC03  & 160.00 & 0.066 & 0.072 & 0.074 \\ 
\midrule
Band & \multicolumn{1}{c}{$\lambda_{\rm lab}$ ({\micron})} & 
\multicolumn{1}{c}{$I$(Obs)} & 
\multicolumn{1}{c}{$I$(1.25\,M$_{\sun}$)} & 
\multicolumn{1}{c}{$I$(1.50\,M$_{\sun}$)}  & 
~~~Band &
\multicolumn{1}{c}{$\lambda_{\rm lab}$ ({\micron})} & 
\multicolumn{1}{c}{$I$(Obs)} & 
\multicolumn{1}{c}{$I$(1.25\,M$_{\sun}$)} & 
\multicolumn{1}{c}{$I$(1.50\,M$_{\sun}$)}  \\ 
\midrule
PAH1 & 8.85 & 52.737 & 52.481 & 50.710 & ~~DUST1 & 22.50 & 323.701 & 292.249 & 286.890 \\ 
PAH2 & 11.30 & 42.644 & 41.541 & 42.593 & ~~DUST2 & 28.50 & 196.829 & 190.268 & 183.637 \\ 
PAH3 & 12.30 & 48.847 & 51.997 & 53.343 &  &  &  &  &  \\ 
\midrule
\end{tabularx}
\begin{tabnote}
Note -- The line and continuum band flux $I$ is normalized to the {\hb} flux $I$({\hb}) = 100. 
The unit of the dust continuum flux density $F_{\nu}$ is Jy. 
PAH1, PAH2, and PAH3 are the PAH band 
fluxes integrated in 
$8.30-9.40$\,{\micron}, 
$11.00-11.60$\,{\micron}, 
and $11.80-12.80$\,{\micron}, respectively.
DUST1 and DUST2 are the dust continuum fluxes integrated 
in $22.00-25.00$\,{\micron} and $27.00-30.00$\,{\micron}, respectively.
\end{tabnote}
\end{table*}

\begin{table}
\caption{The breakdown of the integrated nebular-line and continuum fluxes. \label{T-conti}}
\centering
\begin{tabularx}{\columnwidth}{@{\extracolsep{\fill}}D{-}{-}{-1}c}
\midrule
\multicolumn{1}{c}{Line band ({\micron})} & Flux ({\ergsF})\\ 
\midrule
0.1~-~205             & $3.41\times10^{-12}$\\
\midrule
\multicolumn{1}{c}{Continuum band ({\micron})}  & Flux ({\ergsF})\\ 
\midrule
0.1~-~5.2                  & $7.92\times10^{-12}$\\
5.2~-~36.0                 & $8.22\times10^{-12}$\\
36.0~-~600                 & $9.97\times10^{-12}$\\
\midrule
\multicolumn{1}{c}{Total}  & $6.02\times10^{-11}$\\
\midrule
\end{tabularx}
\end{table}

\end{document}